\documentclass[article,12pt,a4]{JHEP3}
\usepackage[vcentermath,enableskew]{youngtab}
\usepackage{epsfig}
\usepackage{longtable}
\usepackage{amsmath}
\usepackage{amssymb,amsfonts}

\renewcommand{\theequation}{\arabic{section}.\arabic{equation}}
\def\II{\relax{\rm I\kern-.18em I}}
\def\be{\begin{equation}}
\def\ee{\end{equation}}
\def\bs{\begin{subequations}}
\def\es{\end{subequations}}
\def\bc{\begin{center}}
\def\ec{\end{center}}

\newcommand{\een}{\end{subequations}}
\newcommand{\ben}{\begin{subequations}}
\def\beq{\begin{equation}}
\def\eeq{\end{equation}}

\def\hre#1#2{\href{http://arxiv.org/abs/#1/#2}{[ArXiv:#1/#2]}}

\def\cM{{\cal M}}

\def\II{{\cal I}}

\def\MM{{\cal M}}
\def\NN{{\cal N}}
\def\OO{{\cal O}}

\def\bz{\bar z}

\def\s{\sigma}

\def\l{\lambda}
\def\m{\mu}
\def\n{\nu}

\def\a{\alpha}
\def\b{\beta}

\def\sp{\;\;\;,\;\;\;}
\def\r{\rho}

\def\e{\epsilon}

\def\cM{{\cal M}}
\def\cdM{\partial{\cal M}}
\def\bz{{\bar z}}
\def\da{\delta A}

\newcommand\fverb{\setbox\pippobox=\hbox\bgroup\verb}
\newcommand\fverbdo{\egroup\medskip\noindent%
                        \fbox{\unhbox\pippobox}\ }
\newcommand\fverbit{\egroup\item[\fbox{\unhbox\pippobox}]}
\newbox\pippobox

\def\beq{\begin{equation}}
\def\eeq{\end{equation}}
\newcommand{\bea}{\begin{eqnarray}}
\newcommand{\eea}{\end{eqnarray}}

\def\4R{{{}^{(4)}R}}

\def\K5{{\kappa}}
\def\K52{{\kappa^2}}
\def\C{{\cal C}}

\def\pp{\partial}
\def\pa{\partial}

\def\hre#1#2{\href{http://arxiv.org/abs/#1/#2}{[ArXiv:#1/#2]}}

\def\hri#1#2{\href{http://arxiv.org/abs/#1}{[ArXiv:#1]#2}}

\title{Large-N limits of 2d CFTs, Quivers and AdS$_3$ duals}
\author{\href{http://hep.physics.uoc.gr/~kiritsis/}{Elias Kiritsis}$^{a,b}$, \href{mailto:niarchos@physics.uoc.gr}{Vasilis Niarchos}$^a$~\\
~\\
$^a$  \href{http://hep.physics.uoc.gr}{Crete Center for Theoretical Physics},
Department of Physics, University of Crete, 71003 Heraklion, Greece
~\\
$^b$ \href{http://www.apc.univ-paris7.fr/APC_CS/}{Laboratoire APC},
Universit\'e Paris-Diderot Paris 7, CNRS UMR 7164, 10 rue Alice Domon et L\'eonie Duquet, 75205 Paris Cedex 13, France.}

\preprint{CCTP-2010-19}

\abstract{We explore the large-$N$ limits of 2d CFTs, focusing mostly on WZW models and their cosets.
The $SU(N)_k$ theory is parametrized in this limit by a 't Hooft-like coupling. We show a duality
between strong coupling, where the theory is described by almost free fermions, and weak coupling
where the theory is described by bosonic fields by an analysis of spectra and correlators.
The AdS$_3$ dual is described, and several quantitative checks are performed.
Besides the more standard states that should correspond to bulk black holes we
find ground states with large degeneracy that can dominate the standard Cardy
entropy at weak coupling and are expected to correspond to regular horizonless
semiclassical bulk solutions.}


\begin{document}

\section{Introduction and Outlook}

Two dimensional conformal field theories (CFTs) have been studied extensively over
the years. There are many solvable examples that admit a large-$N$ limit. The exact
solvability of these theories in generic regimes of parameter space makes them very
attractive as toy models for questions that are typically very hard for analogous field theories
in higher dimensions. One can compute the exact spectrum and in some cases also exact
correlation functions. In this paper we revisit a special set of two dimensional CFTs: WZW
models (and cosets thereof) based on the $SU(N)$ group. Our interest in these models stems
from the following observations:

\begin{itemize}
\item[$(i)$] On general grounds it is expected that conformal field theories with a
large-$N$ expansion have a dual description in terms of a string theory on AdS.
For the $AdS_3/CFT_2$ correspondence the canonical example is provided
by the symmetric orbifold of $N$ copies of the 2d sigma model with target space
$\MM_4$, which has been argued to be dual to type IIB string theory on
$AdS_3\times S^3\times \MM_4$ (the large-$N$ diagrammatic expansion of
the symmetric orbifold theory has been discussed recently in \cite{Pakman:2009zz}).
In order to gain a deeper insight into the general principles underlying the AdS/CFT
correspondence it is important to extend the list of known examples. Situations involving
solvable CFTs are of obvious interest and hence a natural question is whether it
is possible to identify the string theory duals of well-known exact 2d CFTs. In the
process one hopes to uncover qualitatively new features of the AdS/CFT correspondence.

\item[$(ii)$] We will discuss 2d theories that have interesting analogies/connections to 3d(4d)
field theories. For example, WZW models are well-known to be connected to 3d Chern-Simons
(topological) theories via a bulk-boundary correspondence, albeit not a holographic one
\cite{Jones,elitzur}. Roughly, the quantization of a Chern-Simons (CS) theory with (simple) group
$G$ and coefficient $k\in \mathbb{Z}$ on $\mathbb{R} \times \Sigma$, where $\Sigma$ is a
closed 2d Riemann surface, provides a Hilbert space with states that are in one-to-one
correspondence with the conformal blocks of the $G_k$ current algebra on $\Sigma$.

CS theories coupled to matter also give rise to a generic class of 3d CFTs. Examples with
$\NN=2,3,4,5,6,8$ supersymmetries have been argued to describe the (low-energy) world-volume
theories on multiple M2 branes (see \cite{cs} for a prototype example). M2 branes have two
dimensional boundaries when they end on M5 branes. The theory on these intersections
(self-dual strings) is expected to be a 2d CFT. Recent work \cite{Chu:2009ms} indicates that
this theory involves a WZW model part.

The $AdS_4/CFT_3$ duality for the Chern-Simons-Matter (CSM) theories on M2-branes
predicts that there is a drastic reduction of the degrees of freedom as one moves from weak
to strong 't Hooft coupling. We will observe a qualitatively similar reduction of the degrees
of freedom at strong 't Hooft coupling in two dimensional large-$N$ CFTs.

A good 2d analogue of the (conformal) CS gauge interactions in three dimensions are the
quadratic non-derivative gauge interactions exemplified by two dimensional gauged WZW
models. We will discuss a gauged WZW model that exhibits full level-rank duality \cite{sch}.
This particular duality bears similarities with the Seiberg-like duality of the one-adjoint
$A_{n+1}$ CSM theories in three dimensions \cite{Niarchos:2008jb} and the Seiberg-Kutasov
duality of corresponding one-adjoint SQCD theories in four dimensions \cite{Kutasov:1995ve}.
A relation between level-rank duality in $SU(N)_k$ WZW models and a Seiberg-like duality
in topological $\NN=2$ CS theories was also pointed out in \cite{Aharony:2008gk}. The distinguishing
feature of the level-rank duality that we will discuss here is that it extends to the full theory and
not just to the level of conformal blocks. Accordingly, it bears similarities with 3d (and 4d) Seiberg
duality in non-topological theories.
\end{itemize}

The theories we will be focusing on are the WZW model $SU(N)_k$, \cite{wzw,kz} and its avatars, namely the coset theories, \cite{halpern,gko}, $SU(M+N)_k/SU(N)_k$ and $SU(N)_{k_1}\times SU(N)_{k_2}/SU(N)_{k_1+k_2}$. There are more general cosets, as well as other generalizations of the coset construction, \cite{bos,avc} but we will not consider them here.

We will show that the generating theory, the $SU(N)_{k}$ WZW model, has an interesting 't Hooft
(or Veneziano-like) large-$N$ limit where $N\to\infty$, $k\to \infty$ with ${N\over k}=\l$ fixed.

Moreover, we will show that this theory has two dual descriptions. At weak coupling, $\l\to 0$, the
weakly coupled description is the conventional WZW model, written in terms of a bosonic field $g$
that transforms as a bifundamental under the current algebra symmetry $SU(N)_L\times SU(N)_R$. The theory resembles somewhat the chiral Lagrangian
in four dimensions with important differences: there is no chiral symmetry breaking and the IR theory is conformal.

At strong coupling, $\l\to \infty$, the weakly coupled description is in terms of the IR limit of $N$ copies
of massless Dirac fermions transforming in the fundamental representation of a $U(k)$ gauge group in which the overall $U(1)$ should be thought of as gauged baryon number.
This looks like a conventional gauge theory\footnote{Although the Yang-Mills action is irrelevant in the
IR in two dimensions.} and in this language $k$ is color, $N$ is flavor and $\l={N\over k}$ is the Veneziano ratio. The bosonic field $g$ corresponds to the fermion mass operator in the strongly
coupled regime. In this sense this theory can be thought of as a gauge theory with quarks where although there is confinement, the theory is conformal in the IR and there is no chiral symmetry
breaking. The notion of weak or strong coupling is strictly tied to the definition of $\l$. The weakly coupled WZW theory is strongly coupled from the point of view of the fermionic formulation.

The analogous and more conventional $SU(k)$ gauge theory on the other hand will give rise to the
$U(N)_k$ WZW CFT, with $U(1)_L\times U(1)_R$ the conventional axial and vector baryon number symmetries, both of which are non-anomalous in two dimensions.

This picture is corroborated by studies of the spectrum and four-point functions. This study also gives a concrete example of the non-commutativity of the two limits $N\to \infty$ and $\l\to \infty$. The central charge scales as ${\cal O}(N^2)$, but also depends on $\l$. At strong coupling there is a drastic reduction of the number of degrees of freedom as attested by the value of $c$, not unlike a similar effect in the $AdS_4/CFT_3$ correspondence.

The spectrum is comprised by affine primary ground-states and excitations over them generated by the current modes. Scaling dimensions in the large-$N$ limit take values from ${\cal O}(1)$ to
${\cal O}(N^2)$ for ground states corresponding to representations with about $kN/2$ boxes in the Young tableau. These states have multiplicities that can compete with the Cardy formula.

Such a theory is expected to have an $AdS_3$ dual with $SU(N)_L\times SU(N)_{R}$ symmetry. The closed string sector is expected to be trivial and the dependence on the metric is induced on the boundary by the proper boundary conditions on the open string sector as recognized already in
\cite{Kraus:2006wn}. The open string sector is expected to be realized in a way similar to the one advocated for flavor in higher dimensions. $D_2$ and $\bar D_2$ branes generate the gauge
symmetry which at low energy is realized by two CS actions with couplings of opposite sign.
A direct computation of the effective action for currents is in agreement with the CFT calculation using the WZW model.

The study of scaling dimensions indicates that in the weak coupling limit, the spectrum of ground states can be made to have a large gap from the stringy states. This suggests that in that limit, a ``gravity" description of the physics is possible. In the strong coupling limit all dimensions are of the same order and therefore a stringy bulk description is necessary.
It is in this limit that the physics is described in terms of almost free fermions. The Pauli principle and the existence of a Fermi surface are not visible in the gravity description and a stringy description is necessary in order to accommodate this stringy exclusion principle, very much like in \cite{ms}.

The ground states are generated by a bifundamental field $T$ that should correspond to an  open string stretching between  $D_2$ and $\bar D_2$ and is dual to the generating group field in the WZW model. This is a picture analogous to the one in \cite{tachyon}.
Other ground states in the CFT correspond to multi-particle states. Sources for $T$ correspond to a mass matrix in the fermionic language. They generate a flow that drives the theory to an IR fixed point
equivalent to $SU(N-r)_{k}$ where $r$ the rank of the source of $T$.

The states with scaling dimensions of ${\cal O}(N^2)$ have masses that can be comparable to
$M_{\rm planck}$. We find that they are of the order of $M_{\rm planck}$ at strong coupling and much larger at weak coupling. Therefore, at weak coupling these states could correspond to macroscopic smooth solutions of the bulk theory with associated flavor hair. This hair is responsible for their multiplicity, and in this case and at this energy scale this multiplicity dominates the Cardy entropy.

The cosets are also very interesting CFTs with several possible large-$N$ limits that we analyze.
They also have dual versions, between gauged-WZW models and quiver gauge theories coupled to massless quarks.

The $SU(M+N)_k/SU(N)_k$ theory can be thought of as the following quiver. The gauge group is
$SU(N)\times U(k)$ and the $k(N+M)$ Dirac fermions transform as
$(\Yboxdim8pt\yng(1),\overline{\Yboxdim8pt\yng(1)})$ under $(SU(N),U(k))$ and as $M$ copies of
$(1,\Yboxdim8pt\yng(1))$, having global chiral symmetry $SU(M)_L\times SU(M)_R$.

Finally the $U(k_1+k_2)_N/U(k_1)_N\times U(k_2)_N$ CFT can be thought of as a quiver gauge theory with gauge group $U(k_1)\times U(k_2)\times SU(N)$ and massless quarks transforming as
$(\Yboxdim8pt\yng(1),1,\overline{\Yboxdim8pt\yng(1)})$ and
$(1,\Yboxdim8pt\yng(1),\overline{\Yboxdim8pt\yng(1)})$
under the gauge group. This description automatically explains the level-rank duality symmetry that states that
\be
{U(k_1+k_2)_N\over U(k_1)_N\times U(k_2)_N}\sim {SU(N)_{k_1}\times SU(N)_{k_2}\over SU(N)_{k_1+k_2}}
\ee
The subclass of coset models ${SU(N)_{k_1}\times SU(N)_{1}\over SU(N)_{k_1+1}}$, giving rise to the $W_N$ minimal models, has been analyzed recently in \cite{gg}. Its closed string sector has been argued to correspond to the quantum Hamiltonian reduction $SL(N,C)\to W_N$. Such theories
provide more complex examples of large-$N$ limits but we will only touch upon them in this paper.

There are several issues that remain open in this direction. The first concerns a more organized control
of the spectrum via the partition function. The AdS/CFT correspondence is fundamentally a relation
between partition functions that reads
\beq
\label{introae}
Z_{\rm AdS}=Z_{\rm CFT}
~.
\eeq
In the canonical formulation of the partition function on the torus
\beq
\label{introaf}
Z_{\rm CFT}={\rm Tr}\Big[ e^{2\pi i \tau(L_0-\frac{c}{24})}
e^{-2\pi i \bar \tau (\bar L_0-\frac{\bar c}{24})}\Big]
~.
\eeq
If fermions are present in the theory we also need to specify their periodicities
around the two cycles of the torus. The imaginary part of $\tau$ plays the role
of inverse temperature in the bulk, and the real part is a chemical potential for
angular momentum. There is a reasonably good understanding of this partition function for
the WZW model and the interesting question is whether it can be recast in a way that
makes it interpretable as the partition function on AdS, $Z_{\rm AdS}$. At low
energies the Hilbert space of the gravitational theory comprises of a gas of
particles moving on AdS. At high energies we encounter black holes. There
have been several attempts to make sense of $Z_{\rm AdS}$ as a sum over all saddle
points of the full bulk effective action $I$ (including in principle all string and loop
corrections), $i.e.$ to recast $Z_{\rm AdS}$ as
\beq
\label{introag}
Z_{\rm AdS}(\tau,\bar \tau)=\sum e^{-I}
~.
\eeq
The most concrete realization of this programme is the Farey tail expansion of \cite{Dijkgraaf:2000fq}.
In that case, instead of considering the full partition function, one focuses on a BPS subsector and computes the elliptic genus. Then, one observes that the elliptic genus admits an expansion that is suggestive of a supergravity interpretation in terms of a sum over geometries. It would be interesting to
explore if there is a similar expansion of the {\it full} partition function of the $SU(N)_k$ WZW model.
Practically, a more promising context for this idea is to analyze the elliptic genus of $\NN=2$ gauged WZW models, $e.g.$ $\NN=2$ Kazama-Suzuki models (the supersymmetric Grassmannian coset may be an interesting example).

As we have seen there are many configurations in such CFTs that have macroscopic entropy and are
therefore expected to correspond to smooth bulk solutions carrying flavor hair. It would be interesting to
investigate the existence of such solutions. Moreover, our analysis indicates that at least for the
$SU(N)_k$ theory at strong coupling, such solutions should be thought of as describing a fermionic
ground state (fermi surface) of almost free fermions.

CFTs like the WZW models have a class of features that are interesting to explore in an AdS/CFT setup.
They contain current algebra null vectors that are responsible for the truncation of the spectrum (affine cutoff). It is interesting that although such non-trivial relations are counter-intuitive in the weak coupling limit, they are simply explained in the strong coupling limit where the theory can be described in terms of $kN$ Dirac fermions. {From} the basic formula $g_{a,b}={1\over k}\sum_{i=1}^k\psi^i_a\bar \psi^i_b$
it is then obvious that Fermi statistics forbids symmetrized powers $(g_{ab})^p$ with $p>k$.
The current algebra null vectors are responsible for the existence of ``instanton" corrections to the partition function (namely terms that behave as $e^{-N}$). The study of these effects is interesting as they should match with D-instanton effects in the dual string theory.

A related set of null vectors are the Knizhnik-Zamolodchikov ones, whose content is based on the fact that the stress tensor is quadratic in the currents. They are the key tools in computing the correlation functions. We have seen that the affine-Sugawara construction is an avatar of the proper boundary conditions for the CS theory in 3d. It should be possible to derive the analogue of KZ equations from the bulk.

The bulk description, once developed, will provide a concrete tool for the study of RG flows between different fixed points. A lot is known in 2d CFTs about such flows, but it is expected that the bulk description will provide more efficient tools in this analysis. An intermediate step in this programme is
the understanding of the bulk effective actions for the scalar fields, something that needs further study.
As a byproduct, this approach would allow a more thorough study of the thermodynamics of 2d QFTs.

A related issue is the holographic dual of a $c$-function, a fact that is firmly established for 2d CFTs. It would be interesting to find classes of examples where one could follow the flows between different
fixed points, exploring in such a way the landscape of 2d CFTs. Such large landscapes of flows exist in 2d CFTs, \cite{c}, with the flows and c-function known exactly, although the intermediate non-conformal theories in that case are non-relativistic Hamiltonian theories that flow to standard relativistic CFTs at the fixed points. Recent ideas on ``order" and distance in such a landscape \cite{mrd} may prove useful.

\section{Solvable 2d CFTs}

In two dimensions we have extensive knowledge of a large class of solvable CFTs.
These theories are essentially composed of WZW models based on compact affine Lie
groups, \cite{wzw} and cosets, \cite{halpern,gko}. All of these theories can be solved exactly.
Solvable extensions include some non-compact theories, like Liouville theory, \cite {hd,zamo} and
some semisimple groups \cite{nw,pp}. Larger classes of irrational CFTs were found by
generalizations of the GKO construction, \cite{bos,avc}, but no non-trivial CFT in this class
has been solved so far.

In what follows we will consider the simplest classes of WZW and coset models which possess
large-$N$ limits. But before doing this it is appropriate to make some general comments on
``color" vs ``flavor" degrees of freedom.

We define color as degrees of freedom which are gauged. Gauge fields in two dimensions have
two types of IR dynamics. The standard, namely YM kinetic terms, are irrelevant in the IR, and
never play a role in CFTs. The quadratic gauge terms that play a role are non-propagating ---the
gauged WZW models provide an example where these terms play a defining role. As a result,
pure gauge dynamics in 2d is trivial and the main role of the gauge group is kinematic confinement
and removal of colored degrees of freedom from the spectrum. In this sense, gauge interactions in
two dimensions reduce the number of degrees of freedom the most because of confinement, compared with 3d or 4d gauge dynamics.

On the other hand, we will define flavor degrees of freedom as those degrees of freedom that are
not affected by gauge interactions, in the sense that they are not gauged. For example, a set of $N_f$ free Majorana fermions has a
maximal flavor symmetry $O(N_f)_L\times O(N_f)_R$ that is promoted to an affine current algebra
in two dimensions.

As usual, with the color degrees of freedom being gauged, the relevant operators are gauge singlets  and correspond to closed strings in a dual string theory formulation. The flavor degrees of freedom
being un-gauged correspond in a dual string theory to open strings/D-branes. The fact that 2d pure
gauge theories are trivial, implies the absence of Regge trajectories. Their dynamical role is to
remove matter degrees of freedom. Therefore, the holographic closed string sector dynamics is
typically field theoretic and involves a finite set of states (one of which is the graviton) living on an $AdS_3\times {\cal M}_p$ space. On the other hand, flavor degrees of freedom also have an associated closed string sector: Flavor-singlet operators should be thought of as dual to closed string states.
These are the closed string states that consistently interact with the open string degrees of freedom.

For concreteness, let us now consider some of the examples of interest in this paper.

\subsection{The $SU(N)_k$ WZW model}

This is an interesting case which seems to contain only flavor. The flavor symmetry is
$SU(N)_{L}\times SU(N)_R$ and as it usually happens with non-trivial compact global
symmetries in CFTs, it is extended to a full affine algebra. Note that this not the flavor symmetry
we would obtain from $N$ massless Dirac fermions in two dimensions. That symmetry is larger
and it is $O(2N)_{L}\times O(2N)_R$.

This theory has two parameters, $N$ which is the number of flavors and $k$ that plays the role of the
$\s$-model coupling constant. We will see in the subsequent section that there are several possible large-$N$ limits that can be defined here.

This is a CFT whose spectrum is conveniently represented using current algebra representations. There
are ``ground states'' that coincide with the primary affine representations with spin zero, transforming
as $(R,\bar R)\in SU(N)_{L}\times SU(N)_R$, where $R$ is an integrable representation of the
$SU(N)_k$ affine algebra. All other states are build on the primary states from the action of current operators. They should be thought of as the oscillators of an appropriate open string in AdS$_3$, with the zero mode sector generated by an appropriate CS theory. The closed string states are traces of the flavor degrees of freedom. The stress tensor in particular is composite in the currents and is therefore
not an independent operator. Accordingly, closed string states should presumably be thought of as multiparticle (non-Fock) states of open string states.

As will be explained in section \ref{su(N)k}, the $SU(N)_k$ theory can be thought of as the IR limit of
a theory of $N$ copies of massless Dirac fermions transforming in the fundamental of the gauge
group $U(k)$.

\subsection{The $SU(N)_k/SU(N)_k$ gauged WZW model}

This is the simplest theory that contains $SU(N)$ color but no flavor.
The gauge degrees of freedom remove essentially all states in this theory.
In particular the theory is topological and has central charge $c=0$.
It has a finite number of ground states that are in one-to-one correspondence with all integrable representations of the $SU(N)_k$ affine algebra. In this sense, this Hilbert space can be
thought of as the space of states of a point particle moving on a fuzzy group manifold.

The dual AdS theory is the topological $SU(N)$ Chern-Simons (CS) theory at level $k$.
This theory is topological and has a finite number of states that are also in one-to-one
correspondence with all integrable representations of the $SU(N)_k$ affine algebra, \cite{Jones}.
This is the simplest topological open string theory in 3d. The relevant closed string sector is a
topological string and has no propagating states. {From} the interpretation of the previous section
we gather that the present theory can be thought of as a simple quiver: two gauge groups $U(k)$ and
$U(N)$ and Dirac fermions in the bifundamental. There are no Regge trajectories in this case.

\subsection{The $SU(N+M)_k/SU(N)_k$ gauged WZW model}

This theory has an $SU(N)$ gauge group and therefore $N$ stands for the rank associated to the
color degrees of freedom. The (un-gauged) commuting subgroup of $SU(M+N)$, namely $SU(M)$ should be thought of as a flavor group, while $k$ is a coupling constant.
The central charge is
\beq
c=k\left( \frac{(N+M)^2-1}{k+N+M}-\frac{N^2-1}{k+N}\right)
~.
\eeq
In the 't Hooft limit, $N,k\gg 1$ with the ratio $\lambda=\frac{N}{k}$ fixed, and for $M$ fixed
the central charge becomes to leading order in $M/N$
\beq
c\sim \frac{2+\lambda}{(1+\lambda)^2}NM
~.
\eeq
This is analogous to the quenched limit of 4d gauge theories where the number of flavors is kept finite as the number of colors becomes large.
At weak 't Hooft coupling, $c\sim 2NM$, and the theory looks like a (perturbative) QCD theory with $M$
quarks. Its string theory dual could be identified as an open+closed string theory that arises
by adding $M$ branes in the topological closed string sector of the $M=0$ case.

Another interesting limit of this theory is a Veneziano-type limit where $M/N=m$ is kept fixed and finite.

The coset described here can also be thought of as a quiver. The gauge group is
$SU(N)\times U(k)$ and the $k(N+M)$ Dirac fermions transform as
$(\Yboxdim8pt\yng(1),\overline{\Yboxdim8pt\yng(1)})$ under $(SU(N),U(k))$ and as $M$
copies of $(1,\Yboxdim8pt\yng(1))$, having global chiral symmetry $SU(M)_L\times SU(M)_R$.

\subsection{The $\frac{U(k_1+k_2)_{N}}{U(k_1)_N\times U(k_2)_N}$ gauged WZW model}

This theory can be thought of as a theory with $U(k_1+k_2)$ flavor symmetry whose $U(k_1)\times U(k_2)$ part is gauged. The coupling constant is $N$.
This CFT is dual to the
$$
\frac{SU(N)_{k_1}\times SU(N)_{k_2}}{SU(N)_{k_1+k_2}}$$
coset by level-rank duality, \cite{sch}.
In this dual version of the CFT one starts from an $SU(N)\times SU(N)$ flavor symmetry and then gauges the diagonal subgroup.

{From} the available parameters we can build two independent 't Hooft couplings, $\lambda_i=N/k_i$.
The central charge becomes to leading order in the 't Hooft couplings $c=N^2+subleading$
indicating that this model is similar to a gauge theory coupled to an adjoint scalar. It is therefore interesting to compare it to the $\NN=2$ one-adjoint $A_{n+1}$ CSM theories of
\cite{Niarchos:2008jb}. Some parallels between these theories are:
\begin{itemize}
\item[$(a)$] Both are controlled by three discrete parameters. In the CSM case,
these three parameters are: $k$ the level of the CS interaction, $N$ the rank of the
$U(N)$ gauge group, and $n+1$ the power of the single-trace operator
${\rm Tr} X^{n+1}$ that appears in the action as a superpotential deformation.
\item[$(b)$] A crucial effect of the superpotential deformation in the CSM theory is that it
truncates the chiral ring. The levels $k_i$ play an analogous role in the WZW model truncating
the spectrum.
\item[$(c)$] Both theories exhibit a non-trivial duality. The $U(N)_k$ $A_{n+1}$ CSM theory
is Seiberg-dual to the $U(nk-N)_k$ $A_{n+1}$ CSM theory.
The $\frac{SU(N)_{k_1}\times SU(N)_{k_2}}{SU(N)_{k_1+k_2}}$ gauged WZW model is dual,
by level-rank duality, to the ${SU(k_1+k_2)_N\over SU(k_1)_N\times SU(k_2)_N\times U(1)}$ gauged
WZW model.
\end{itemize}

We can also think of the $\frac{U(k_1+k_2)_{N}}{U(k_1)_N\times U(k_2)_N}$ theory
as a quiver gauge theory with gauge group $U(k_1)\times U(k_2)\times U(N)$ and massless quarks transforming as
$(\Yboxdim8pt\yng(1),1,\overline{\Yboxdim8pt\yng(1)})$ and
$(1,\Yboxdim8pt\yng(1),\overline{\Yboxdim8pt\yng(1)})$
under the gauge group. This description automatically explains the level-rank duality symmetry.

\section{On large-$N$ limits}

There are several large-$N$ limits that are possible in the CFTs we have mentioned above.
They have been discussed in different contexts in the literature and we will go through them for comparison. As the CFTs in question are solvable, we will be able to characterize explicitly the
nature of each of these limits.

\subsection{The 't Hooft large-$N$ limit\label{hooft}}

The characteristic feature of the 't Hooft limit is that the coupling constant is rescaled so that it compensates for the increase of degrees of freedom. Another characteristic is that for adjoint
theories the normalized $n$-point functions behave as $N^{1-{n\over 2}}$.
This implies in particular that the central charge $c\sim {\cal O}(N^2)$.
For the $SU(N)_k$ theory the 't Hooft limit implies $N\to \infty$, $k\to\infty$ with
\be
\lambda={N\over k}
\label{a19}\ee
kept fixed, \cite{product,gg}. When $\lambda\ll 1$ we are in a perturbative regime. In this regime the
$\alpha'$-perturbation theory is applicable. Since in this limit, $k\gg N$, the affine cutoff \cite{gw} is
not visible when we consider representations with $\sim {\cal O}(N)$ columns in the Young tableau or less. Therefore, the fusion algebra of such low-lying representations is ``perturbative": it coincides with the classical Glebsch-Gordan decomposition.

In the opposite limit, $\l\gg 1$, we are in the strong coupling regime. The $\s$-model semiclassical
expansion breaks down and since $k\ll N$ the affine cutoff is felt at relatively low representations.
This implies algebraic relations between primary fields (the vanishing of fields with spin higher than
$k$) well before the distinction between single-trace and double-trace operators sets in.

The central charge can be written in this limit as
\be
c={N^2\over 1+\lambda}+ {\cal O}(1)
\label{a20}\ee
and it is indeed ${\cal O}(N^2)$ as advertised.
It does remain so at weak 't Hooft coupling but at strong coupling
\be
c\sim {N^2\over \lambda}+{\cal O}\left({1\over \l^2}\right)\sim kN+{\cal O}\left({1\over \l^2}\right)
\label{a21}\ee
is parametrically smaller than $N^2$ and behaves as ${\cal O}(N)$ for finite $k$. As the central charge is a quantum measure of the number of degrees of freedom, this indicates that there is a drastic reduction
of degrees of freedom at strong 't Hooft coupling mimicking a similar situation predicted by the
$AdS_4/CFT_3$ correspondence for three dimensional conformal field theories.
An important difference is that here this reduction is explicitly calculable.\footnote{Recently, an
analogous result in three dimensions was computed from a reduced matrix model, \cite{dmp}.}
Another difference is that the reduction observed here is by a factor of $N$ while in three dimensions
it is by a factor of $\sqrt{N}$.\footnote{An analogous analysis in the $AdS_7/CFT_6$ correspondence
for M5 branes indicates that the $CFT_6$ at strong 't Hooft coupling has more rather than less degrees of freedom from an equivalent weakly coupled theory.}

Focusing at the conformal dimensions with an ${\cal O}(1)$ number of boxes of the Young tableau,
we obtain (see appendix \ref{casimir})
\be
\Delta_R={\l\over 1+\l}\Delta_R(\infty)+{\cal O}\left({1\over N}\right)
~.
\label{a22}\ee
To leading order at $1/N$ and at weak coupling they all asymptote to zero
\be
\Delta_R=\l\Delta_R(\infty)+{\cal O}\left(\l^2\right)~,
\label{a23}\ee
in agreement with the fact that in the classical theory all primary operators have vanishing scaling dimension.

In general, in the 't Hooft limit  the dimensions of primaries are
\be
\label{a25a}
\Delta_{R}={\l\over 2(\l+1)}\sum_{i=1}^{k} m_i +{\cal O}\left({1\over N}\right)
\ee
where $m_i$ are the Dynkin indices  provided the sums are ${\cal O}(1)$. Otherwise, the full formula
(\ref{d}) should be used.
All $m_i$ take values $0\leq m_i\leq N/2$. In all cases, $\sum_{i=1}^{k} m_i$
is the total number of boxes in the Young tableau of the associated representation.

At strong 't Hooft coupling on the other hand they asymptote to half-integers
\be
 \Delta_R=\Delta_R(\infty)+{\cal O}\left({1\over \l}\right)~.
\label{a24}\ee
Note that this is the same spectrum as in the naive large-$N$ limit discussed in section \ref{naive}.

The maximal dimension is obtained when $m_i=N/2$ $\forall ~i$. In that case,
\be
\label{a26a}
C_2={kN(k+N)\over 8}\sp \Delta_{max}={\l\over 2(\l+1)}{kN\over 2}={N^2\over 8\lambda}~.
\ee
As we will later see, this state is expected to correspond to a regular horizonless semiclassical
bulk solution.

On the other hand, for the maximal symmetric tensor, $m_i=1$ $\forall ~i\leq k$,
\be
\label{a27a}
C_2={k(k+N)\over 2}\sp \Delta_{sym}={N\over 2\lambda}+\cdots~.
\ee
For the maximal antisymmetric representation, $m_1=N/2$ and all other $m_i$'s zero,
\be
\label{a28a}
C_2={N^2\over 8}\sp \Delta_{a}={\l\over 8(\l+1)}N+\cdots
~.
\ee

\subsection{The simple large-$N$ limit\label{naive}}

Another possibility is to take $N\to \infty$ while keeping $k$ fixed.
This limit has been studied previously in specific examples, in \cite{bk,bk2}, in order to produce representations of the $W_{\infty}$ algebra. In this limit
\be
c\simeq kN+{\cal O}(1)
\label{a6} \ee
and in this sense it looks like the theory is reducing in this case to a vectorial large-$N$ theory.
The large-$N$ limit of the dimensions of primary fields gives
\be
\Delta_R\simeq \Delta_R(\infty)+{\cal O}\left({1\over N}\right)~.
\label{a7} \ee
Therefore the primary field dimensions become half-integers which hints at a free-fermionic
formulation in terms of $2kN$ free fermions (as also suggested by the central charge). This is
indeed true as analyzed in appendix \ref{free}.

The formula (\ref{a7}) applies to representations with an ${\cal O}(1)$ number of boxes in the Young-tableau. However, if one considers representations with ${\cal O}(N)$ boxes then things are different.
For example, the antisymmetric representation with ${N\over 2}+m$ boxes has dimension
\be
\Delta_{A_{{N\over 2}+m}}={N+1\over 8}-{m^2\over 2N}+{\cal O}\left({1\over N^2}\right)~.
\label{a8} \ee
On the other hand, the maximal symmetric representation can only have $k$ boxes because of
the affine cutoff. We therefore have
\be
\Delta_{S_k}={k\over 2}+{\cal O}\left({1\over N}\right)~.
\ee
Hence, we have dimensions scaling as ${\cal O}\left({ N}\right)$ and dimensions scaling as
${\cal O}\left({1}\right)$.

In coset theories primary field dimensions can also be of the order of ${\cal O}\left({1\over N}\right)$
as shown in \cite{bk,gg}. Consider the coset
\be
CFT\equiv {SU(N)_{k_1}\times SU(N)_{k_2}\over SU(N)_{k_1+k_2}}\sp c={k_1k_2(k_1+k_2+2N)(N^2-1)\over (k_1+N)(k_2+N)(k_1+k_2+N)}
\label{a9} \ee
In the naive large-$N$ limit we may rewrite the central charge as
\be
c=2k_1k_2+{\cal O}\left({1\over N}\right)
\label{a10} \ee
which is finite in the large-$N$ limit.

There is an interesting symmetry in this theory, stemming from level-rank duality, that indicates that
this CFT is equivalent to a dual one\footnote{This has been explicitly checked in the associated supersymmetric models, \cite{sch}.}
\be
CFT\sim \widetilde{CFT}\equiv {SU(k_1+k_2)_N\over SU(k_1)_N\times SU(k_2)_N\times U(1)}
\label{a11} \ee

The dimensions of the coset are associated with three representations: $R_1\in G_1$, $R_2\in G_2$ and $R_3\in R_1\otimes R_2$.
Since
\be
\Delta_{R_1,R_2;R_3}=\Delta_{R_1}+\Delta_{R_2}-\Delta_{R_3}+{\rm integer}
\label{a12} \ee
we obtain in the large-$N$ limit
\be
\Delta_{R_1,R_2;R_3}=\Delta_{R_1}(\infty)+\Delta_{R_2}(\infty)
-\Delta_{R_3}(\infty)+{\cal O}\left({1\over N}\right)~.
\label{a13} \ee
The `+integer' on the right hand side of \eqref{a12} denotes a possible contribution that
appears in some cases
because the coset primary is not the top state of the representation but one of the descendants.
{}For representations with an ${\cal O}(1)$ number of boxes in the Young tableau
this dimension is of order ${\cal O}\left({1\over N}\right)$.
For example, consider the case $R_1=A_{m_1}$, $R_2=A_{m_2}$, $R_3=A_{m_1+m_2}$.
We obtain
\be
\Delta_{A_{m_1},A_{m_2};A_{m_1+m_2}}={m_1m_2\over N}+{\cal O}\left({1\over N^2}\right)~.
\label{a14}\ee
We will still have also dimensions that scale like ${\cal O}\left({N}\right)$.
For example,
\be
\Delta_{A_{{N\over 2}-m_1},A_{{N\over 2}-m_2};A_{N-m_1-m_2}}={N+1\over 4}-{m_1+m_2\over 2}+{\cal O}\left({1\over N}\right)~.
\label{a15}\ee

Therefore in such CFTs, there are primary operators in class A with dimensions   ${\cal O}\left({N}\right)$,
operators in class B with dimensions ${\cal O}\left({1}\right)$ and operators in class C with ${\cal O}\left({1\over N}\right)$.
Moreover, the maximum dimension is obtained with $m_1=m_2=\cdots =m_{k-1}={N\over 2}$ with
\be
\Delta_m\simeq {kN\over 8}+\cdots~.
\ee

It was shown in \cite{bk} that one can construct a class of operators of dimension
${\cal O}\left({1}\right)$ out of operators of the class C. The construction involves operators as in
(\ref{a14}) and the limit $m_1=q_1\sqrt{N}$, $m_2=q_2\sqrt{N}$ with $q_1,q_2$ fixed.
Such operators were shown to have abelian OPEs and generate the analogue of the discrete series operators in pp-wave CFTs as shown in \cite{pp}.

Note that this large-$N$ limit has a dual version in the $\widetilde {\rm CFT}$ (\ref{a11}) as a weak coupling limit where all current algebra levels go to infinity. Therefore the $\s$-model is a flat space to leading order with $2k_1k_2$ dimensions, and the states in this theory, correspond to the
class B operators as well as the class C operators that can be made to have ${\cal O}(1)$ dimensions as explained earlier.

Note also that the $U(N)_1$ theory is equivalent to a collection of $2N$ free fermions $\sim O(2N)_1$, as described in appendix \ref{free}. The only integrable representations in this case are the
antisymmetric ones and are constructed from products of fermions.
The $SU(N)_1$ is obtained from the free fermion theory by coseting the overall $U(1)$.

\subsection{The BMN large-$N$ limit\label{bmn}}

This is a large-$N$ limit with a tuning that keeps some dimensions finite, \cite{olive,pp}.
We will consider the following example: $SU(N)_k\times U(1)^{N-1}$ where the $U(1)$'s are
time-like and have level $2N$. The BMN limit is a large-$N$ limit at fixed $k$ that ties together specific combinations of $U(1)$ and $SU(N)$ representations. This generates a contraction of the group to a non-semi-simple one where one linear combination of the $U(1)$'s and $SU(N)$ Cartan generators
becomes a set of $N$-1 central currents, the other linear combination becomes a set of $N$-1 rotation operators, and all raising and lowering operators become transverse pp-wave operators.

The dimension of a generic primary is
\be
\Delta_{\vec q,R}=-{\vec q^2\over 2N}+{\C_2(R)\over k+N}~.
\label{a16}\ee
Consider a representation of type A with $x_iN+\xi_i$ boxes in the $i$-th column, with $0<x_i<1$, $i=1,2,\cdots,k$ and $\xi_i$ of order ${\cal O}(1)$.
This representation has a dimension of order ${\cal O}(N)$. We also pick the $U(1)$ charges $q_i$
so that they cancel the ${\cal O}(N)$ piece of the previous dimension
\be
q_i=N\sqrt{x_i(1-x_i)}~.
\label{a17}\ee
In the large-$N$ limit
\be
\Delta=\sum_{i=1}^k\xi_i(1-2x_i)+(1-k)x_i(1-x_i)~.
\label{a18}\ee

Such states provide highest weight or lowest weight representations. If the CFT has operators of
type C, then their continuous limit with $U(1)$ charges of order
${\cal O}(1)$ provides continuous series representations \cite{pp}.

\section{The $SU(N)_k$ WZW model\label{su(N)k}}

This is a prototypical unitary CFT, realizing a current algebra that depends on two natural numbers,
$N,k$. The global symmetry $SU(N)_L\times SU(N)_R$ is enhanced to the full affine left-moving and right-moving algebra $SU(N)_k$.

It can appear as an IR fixed point in many CFTs, including the $SU(N)$ chiral model modified by the addition of a WZ term, \cite{wzw}. It can also appear as the IR fixed point of 2-dimensional massless QCD with gauge group $U(k)$ and $N$ Dirac flavors of quarks \cite{2d,cobi}, as explained in appendix
\ref{free}. In such a description the YM action is becoming irrelevant in the IR and the theory flows to the $U(kN)_1/U(k)_N$ coset that is equivalent to the $SU(N)_k$ CFT. Therefore $k$ can be identified as
the number of colors and $N$ as the number of massless quark flavors.

We will denote the ratio of flavor to color numbers as
\be
\l={N\over k}
\label{1}\ee
and we will call it the 't Hooft coupling, although from the fermionic point of view, it is the Veneziano ratio $N_f\over N_c$. In the $\sigma$-model picture this ratio does look more like the conventional 't Hooft coupling.

In the gauge theory picture the global flavor symmetry is $SU(N)_L\times SU(N)_R$ and is manifest in the theory. This is an example of a confining gauge theory without chiral symmetry breaking. The reason that such a theory can still be a CFT is that in two dimensions non-abelian gauge fields carry no propagating degrees of freedom and confinement is essentially kinematical. It can be implemented by the (generalized) Gauss law of the appropriate current algebra on the states of the un-gauged theory.

Denoting the currents of the $SU(N)_k$ WZW model as $J^a(z)$, $\bar J^a(\bar z)$ we have the OPEs
\beq
\label{specaa}
J^a(z)J^b(w) \sim \frac{k\delta_{ab}}{(z-w)^2}+\sum_c i f_{abc} \frac{J^c(w)}{z-w}
\eeq
with a corresponding expression for the right-moving currents $\bar J^a$ (which we will systematically omit).

The stress-energy tensor satisfies the affine Sugawara construction
\beq
\label{specab}
T(z)=\frac{1}{2(k+N)}\sum_a (J^a J^a)(z)
~.
\eeq
 The central charge in this theory is given by
 \be
 c={k(N^2-1)\over k+N}={(N^2-1)\over \l+1}\sim {\cal O}(N^2)\sim {\cal O}(k^2)~.
\label{a1} \ee

In the large 't Hooft limit, $\l\gg 1$, $c\simeq {N^2\over \l}=kN$.
One can observe the same phenomenon that has been observed in 3d CFTs dual to M2 brane geometries. At strong 't Hooft coupling there is a reduction of the number of degrees of freedom;
in this case, from ${\cal O}(N^2)$ to ${\cal O}(N)$. As shown in appendix \ref{free}, in this regime of
$N\gg k$, the theory is approximately described by $kN$ free massless Dirac fermions and the
YM interaction can be treated perturbatively.
In the opposite regime, $\l\ll 1$ or $N\ll k$, the theory is well described by the (weakly coupled) WZW theory, describing string propagation over a large volume group manifold.

We now proceed to analyze the spectrum and conformal dimensions.
The spectrum is composed of affine (spinless) primary states transforming in the $(R,\bar R)$ integrable representations of the global  $SU(N)_L\times SU(N)_R$ group, with one copy per representation.
On top of these primary states the whole affine representation is built by acting with the lowering operators of the current algebra, the negative $J_{-n}^a$ current modes. Were it not for the existence of non-trivial current algebra null vectors, the multiplicity of the states inside representations would be the same as that of the related $U(1)^{N^2-1}$ free theory.
The current algebra null vectors provide non-perturbative effects in $k$, and disappear as $k\to \infty$.

The (left-moving) conformal dimensions for the affine primary fields of the $SU(N)_k$  theory, transforming in the $R$ irreducible unitary representation of the $SU(N)$ algebra are given by
\be
\Delta_R={C_2(R)\over k+N}
\label{a2} \ee
where $C_2(R)$ is the quadratic Casimir, defined and analyzed in appendix \ref{casimir}. To find the large-$N$ limits we must analyze the scaling of the quadratic Casimir for $SU(N)$ representations,
with the result
\be
C_2(R)\simeq N\Delta_R(\infty)+{\cal O}\left({1}\right)
\label{a5} \ee
in the $N\to \infty$ limit for representations with an ${\cal O}(1)$ number of boxes in their Young
tableau.

We can give a general formula for the quadratic Casimir $C_2(R)=C_2(\bar R)$, corresponding to a Young tableau with $m_1$ boxes in the first column, $m_2$ boxes in the second column etc, with the $m_i$ ordered, $m_1\geq m_2\geq m_3\cdots$. We use the prescription that the proper Young
tableau is the one with the minimum number of boxes. In this notation the Casimir is
\be
C_2(m_1,m_2,\cdots,m_n)={(\sum_{i=1}^nm_i)N^2-((\sum_{i=1}^n m_i^2)-\sum_{i=1}^n (2i-1)m_i)N-(\sum_{i=1}^n m_i)^2\over 2 N}~.
\label{d}\ee
For $m_i\sim \OO(1)$ and $n\sim {\cal O}(1)$ we obtain
\be
\Delta_R(\infty)={1\over 2}(\sum_{i}m_i)
\label{3}\ee
which are half-integers. In particular, this number is the total number of boxes in the Young tableau, divided by two.

We finally obtain
\be
\Delta_R\to {\l\over \l+1}\Delta_R({\infty})+{\cal O}\left({1\over N}\right)~.
\label{4}\ee
In the strong coupling limit $\lambda\to \infty$, the dimensions are half-integers,
reflecting the fact that the theory asymptotes to a theory of $kN$ free massless fermions
(see appendix \ref{free} and the next section). In that case one can think of a rectangular box of
dimensions $N\times k$ divided into $kN$ compartments (boxes). A choice of ``occupied" boxes defines
a Young tableau and therefore an integrable representation $R$ of $SU(N)_k$.
Not surprisingly, the set of ``ground-states" of the theory, namely the primary fields, is in one-to-one correspondence with filling some of the boxes using the $kN$ fermions while abiding to the Pauli
principle.

We will now consider representations with large scaling dimensions.
The maximum scaling dimension is given for representations around the ``half-box",
\be
m_i={N\over 2}+n_i\sp n_i\sim {\cal O}(1)
\label{5}\ee
with
\be
C_2={Nk(N+k)\over 8}-{\sum_{i=1}^k n_i^2+\sum_{i=1}^k (k+1-2i)n_i+{1\over N}(\sum_{i=1}^k n_i)^2\over 2}
\label{6}\ee
with scaling dimension
\be
\Delta={N^2\over 8\lambda}-{\lambda\over 2(\lambda+1)}\left[{1\over N}\sum_{i=1}^k n_i^2+
{1\over N}\sum_{i=1}^k (k+1-2i)n_i+{1\over N^2}(\sum_{i=1}^k n_i)^2\right]~.
\label{7}\ee
Note that the scaling dimension is of order ${\cal O}(N^2)$, and is 1/8 times the central charge
in (\ref{a1}).

The spectrum of the theory is therefore consisting of ``ground states'' associated with the affine primary fields and ``stringy excitations" associated with the affine descendants, generating Regge trajectories
on top of the ground-states.

The ground state transforming under the representation $(R,\bar R)$ of the global group $SU(N)_L\times SU(N)_R$ has multiplicity $D(R)^2$ where $D(R)$ is the dimension of the associated representations. The dimensions start at ${\cal O}(1)$ but the multiplicity they generate can be substantially higher.

For example, the maximal representation corresponding to the ``half-box" in (\ref{5}) with $n_i=0$ has dimension
\be
D={\prod_{i=1}^{N/2}{(N+k-i)!\over (N-i)!}\over \prod_{i=1}^{k}{\left({N\over 2}+i-1\right)!\over (i-1)!}}~.
\label{9}\ee
The logarithm of the multiplicity of this ground state is derived in appendix \ref{casimir} as
\be
\log D^2=\left[4(\l+1)\log 2-\l^2\log \l+2(\l+1)^2\log(\l+1)-(\l+2)^2\log(\l+2)\right]{N^2\over 2\l^2}+
\label{10}\ee
$$
-{N\over 2}\log N+\left[2\log(\l+2)-4(\l+1)\log (\l+1)+2(\l-1)\log \l+2\l \log 2\right]{N\over 2\l}
-
$$
$$
-{1\over 6}\log{\l+2\over \l}+{\cal O}(N^{-1})~.
$$

At strong coupling, $\l\gg 1$,
\be
\log D^2\simeq (2\l\log 2-\log \l+\cdots){N^2\over \l^2}+{\cal O}(N\log N)\sim {\cal O}\left({N^2\over \l}\right)\simeq 2c\log 2 +\cdots
\label{11}\ee
while at weak coupling, $\l\ll 1$,
\be
\log D^2\simeq \left[{\log{1\over \l}\over 2}+{\cal O}(\l^2)\right]N^2+{\cal O}(N\log N)\sim {\cal O}\left(N^2\log{1\over \l}\right)\simeq c \log{1\over \l}+\cdots~.
\label{12}\ee

We should also consider states whose scaling dimensions scale as ${\cal O}(N)$ in the large-$N$
limit. In section \ref{hooft} we have mentioned two such representations; the maximal symmetrized representation with all Dynkin indices $m_{0<i<k+1}=1$, and the ``maximal" antisymmetrized representation, $m_1={N\over 2}$ and all other $m_i$'s zero.
The scaling dimension of the maximal symmetric tensor is
\be
\label{bis1}
C_2={k(k+N)\over 2}\sp \Delta_{sym}={N\over 2\lambda}+\cdots
\ee
while for the antisymmetric one
\be
\label{bis2}
C_2={N^2\over 8}\sp \Delta_{a}={\l\over 8(\l+1)}N+\cdots~.
\ee
Note that $\Delta_{sym}$ vanishes in the strong coupling limit. This is a consequence of
Fermi-statistics, as this representation is built out of fermions as $\prod_{i=1}^k\psi^{a_i}_i\times cc$,
and therefore one cannot obtain a symmetric object in the flavor indices, $a_i$.
On the other hand, $\Delta_a\to {N\over 8}$ in the strongly coupled limit as in this case a similar state, can be achieved in terms of free fermions. Such states resemble baryons, and indeed their
``masses" are of order $N$. Of course there are many more representations of this type, beyond
the ones discussed above.

Note that the $N$ dependence of scaling dimensions that arises in (\ref{a2}), (\ref{d}) contains contributions from tree-level, disk-level and one-loop. On the other hand, the central charge has only
a single one-loop contribution. The $\l$ dependence although simple indicates the presence of a full perturbative series of corrections both at small and large $\l$.

We have seen that the Hilbert space decomposes into a finite number of unitary irreducible representations of the current algebra. For each representation there is a ground state, transforming in the $(R,\bar R)$ of $SU(N)_L\times SU(N)_R$ corresponding to the primary field, and the rest of the states are generated from the primary field by the action of the current oscillators.
The different primary operators can be thought of as products of the basic WZW field $g_{a,b}(z,\bar z)$.
Using appropriate normal ordering and symmetrizations or antisymmetrizations of the indices we may construct any integrable primary of the algebra.
This fact will be important when we analyze the AdS$_3$ dual.

\subsection{The four-point function \label{4pp}}

An important observable in conformal field theory are correlation functions.

We will analyze the large-$N$ limits of the four-point function of the basic primary fields of the
$SU(N)_k$ WZW model, namely the fundamental $g_{a,b}(z,\bar z)$ and its conjugate
$g^{-1}_{b,a}(z,\bar z)$ following \cite{kz} and then study its large-$N$ limit.

The four-point function has been calculated solving the KZ equation and reads
\be
G(x,\bar x)\equiv\langle g_{a_1,b_1}(\infty)g^{-1}_{b_2,a_2}(1)g_{a_3,b_3}
(x,\bar x)g^{-1}_{b_4,a_4}(0)\rangle=\sum_{A,B=1}^2 I^A \bar I^B~G_{AB}(x,\bar x)
\label{14} \ee
\be
I^1=\delta_{a_1,a_2}\delta_{a_3,a_4}\sp   \bar I^1=\delta_{b_1,b_2}\delta_{b_3,b_4}
\sp I^2=\delta_{a_1,a_4}\delta_{a_2,a_3}\sp   \bar I^2=\delta_{b_2,b_4}\delta_{b_1,b_3}
\label{15}\ee
with
\be
G_{AB}(x,\bar x)={\cal F}^{(1)}_A(x){\cal F}^{(1)}_B(\bar x)+h~{\cal F}^{(2)}_A(x){\cal F}^{(2)}_B(\bar x)
~.
\label{16}\ee

There are two group invariants corresponding to the two representations, the singlet and the adjoint appearing in the product of fundamental with an anti-fundamental. Taking into account the left-moving and right-moving group structure, the total number of invariants is four and they appear in (\ref{15}). $h$ in (\ref{16}) is the only non-trivial quantum Glebsch-Gordan (OPE) coefficient coupling a fundamental,
an anti-fundamental and the adjoint.

A detailed analysis is presented in appendix \ref{4p}. In what follows we summarize the main
results.

The associated conformal blocks are
\be
{\cal F}^{(1)}_1(x)=x^{-2\Delta_{\Yboxdim4pt\yng(1)}}(1-x)^{\Delta_A-2\Delta_{\Yboxdim4pt\yng(1)}}
~F\left(-{1\over 2\kappa},{1\over 2\kappa};1+{N\over 2\kappa},x\right)~,
\label{17}\ee
\be
{\cal F}^{(1)}_2(x)=-{x^{1-2\Delta_{\Yboxdim4pt\yng(1)}}(1-x)^{\Delta_A-2\Delta_{\Yboxdim4pt\yng(1)}}
\over 2\kappa+N}
~F\left(1-{1\over 2\kappa},1+{1\over 2\kappa};2+{N\over 2\kappa},x\right)~,
\label{18}\ee
\be
 {\cal F}^{(2)}_1(x)=x^{\Delta_A-2\Delta_{\Yboxdim4pt\yng(1)}}(1-x)^{\Delta_A-2\Delta_{\Yboxdim4pt\yng(1)}}
 ~F\left(-{N-1\over 2\kappa},-{N+1\over 2\kappa};1-{N\over 2\kappa},x\right)~,
\label{19} \ee
\be
{\cal F}^{(2)}_2(x)=-Nx^{\Delta_A-2\Delta_{\Yboxdim4pt\yng(1)}}(1-x)^{\Delta_A-2\Delta_{\Yboxdim4pt\yng(1)}}
 ~F\left(-{N-1\over 2\kappa},-{N+1\over 2\kappa};-{N\over 2\kappa},x\right)~,
\label{20} \ee
where $F$ is the hypergeometric function and
\be
\Delta_{\Yboxdim4pt\yng(1)}={N^2-1\over 2N(N+k)}\sp \Delta_A={N\over N+k}\sp 2\kappa=-N-k
\label{21}\ee
and
\be
h={1\over N^2}{\Gamma\left[{N-1\over N+k}\right]
\Gamma\left[{N+1\over N+k}\right]\Gamma^2\left[{k\over N+k}\right]
\over \Gamma\left[{k+1\over N+k}\right] \Gamma\left[{k-1\over N+k}\right] \Gamma^2\left[{N\over N+k}\right]}~.
\label{22} \ee

In the 't Hooft limit we obtain
\be
G_{11}=|x|^{-{2\l\over 1+\l}}+{\cal O}\left({1\over N^2}\right)\sp  G_{22}=|1-x|^{-{2\l\over 1+\l}}+
{\cal O}\left({1\over N^2}\right)~,
\label{23} \ee
\be
G_{12}= {\lambda \bar x\over N}|x|^{-{2\l\over 1+\l}}~~F\left(1,1;{2+\l\over 1+\l},\bar x\right)
-{(1-x)\over N}|1-x|^{-{2\l\over 1+\l}}~F\left(1,1;1+{\l\over 1+\l},x\right)+{\cal O}\left({1\over N^2}\right)~.
\label{24}\ee

Note that in the symmetric channels $G_{11}$ and $G_{22}$ are powerlike at leading order in $1/N$ suggesting the existence of a single block. In the asymmetric channel $G_{12}$ there is non-trivial
structure at subleading order in $1/N$.

Finally in the strong coupling limit $\l\to \infty$ the result simplifies to
\be
G_{11}={1\over |x|^{{2}}}+\cdots\sp  G_{22}={1\over |1-x|^{{2}}}+\cdots\sp
G_{12}= {\lambda\over N}{1\over x(1-\bar x)}+\cdots~.
\label{a40}\ee
This is indeed compatible with the claim that the theory is described by free fermions in that limit.
In particular, the properly normalized fundamental field of the WZW can be written in the strong
coupling limit as
\be
g_{ab}(z,\bar z)={1\over k}\sum_{i=1}^k\psi^a_i(z)\tilde \psi^b_i(\bar z)
\label{a41} \ee
where the tilde indicates right-movers. {From} (\ref{a41}) the 4-point function in (\ref{a40}) follows.

\subsection{Non-commutativity of large-$N$ and large-$\l$ limits.}

An interesting question in any large-$N$ theory is the commutativity of the large-$N$ and the
large 't Hooft coupling limit. In the conventional definition we first take the large-$N$ limit, and
then let $\l$ become large. In our example we can study these limits explicitly and we will show
that they do not commute. More details can be found in appendix \ref{4p}.

The particular observable to study is the OPE coefficient in (\ref{22}).
This OPE coefficient has a double expansion
\be
h=\sum_{n=1}^{\infty}\sum_{m=-2}^{\infty}{W_{n,m}\over N^{2n}\l^m}~.
\label{a27} \ee

If we first take the large-$N$ limit while keeping $\l$ fixed we re-organize the double expansion as
\be
\lim_{N\to\infty}~h=\sum_{n=1}^{\infty} {Z_{n}(\l)\over N^{2n}}~.
\label{a28} \ee
Next we take the large-$\l$ limit. The functions $Z_n$ have the following behavior in this limit
\be
\lim_{\l\to\infty} Z_2=-\l^2+{\cal O}\left({1\over \l}\right)~,
\label{a30} \ee
\be
\lim_{\l\to\infty} Z_3=2\l\psi''(1)-6\psi''(1)+{\cal O}\left({1\over \l}\right)~,
\label{a31} \ee
from which we read in the range $1\leq n\leq 3$, $-2\leq m\leq 0$
\be
W_{1,0}=1\sp W_{2,-2}=-1\sp W_{3,-1}=2\psi''(1)\sp W_{3,0}=-6\psi''(1)
\label{aa32} \ee
all other being zero in that range. $\psi(x)=\frac{d \ln \Gamma(x)}{dx}$ is the standard digamma function.

We will now take the opposite sequence of limits; the large-$\l$ limit coming first. We rewrite the double expansion as
\be
\lim_{\l\to\infty}~h=\sum_{n=-2}^{\infty}H_{n}(N)\l^{-n}
\label{a33} \ee
and subsequently take the large-$N$ limit of the functions $H_n$. We obtain
\be
\lim_{N\to\infty} H_{-1}={2\psi''(1)\over N^6}+{\cal O}(N^{-8})
\label{a36} \ee
\be
\lim_{N\to\infty} H_{0}={1-4\gamma_E+{\pi^2\over 3}\over N^4}+{15\pi^2+\pi^4-180\psi''(1)\over 45N^6}+{\cal O}(N^{-8})
\label{a37} \ee
from which we deduce
\be
W_{2,-2}=-1\sp W_{3,-1}=2\psi''(1)\sp W_{2,0}=1-4\gamma_E+{\pi^2\over 3}\sp W_{3,0}={15\pi^2+\pi^4-180\psi''(1)\over 45}
\label{aa38} \ee
while the rest are zero.

Comparing the expressions (\ref{aa32}) and (\ref{aa38}) we observe that the two limits do not commute.

\subsection{The effective action for sources\label{ea}}

We will derive here the effective action of the WZW theory once we couple the currents to sources. The WZW action is given by
\be
I(g)={1\over 16\pi}\int d^2\xi Tr[\partial_a g\partial^a g^{-1}]+\Gamma(g)
\sp\Gamma(g)={i\over 24\pi}\int d^3\xi Tr[g^{-1}\partial_a g g^{-1}\partial_b g g^{-1}\partial_c g]\e^{abc}
\label{e1}\ee
where the second integral is over a 3d manifold with our two dimensional space as its boundary.
The action satisfies the Polyakov-Wiegmann relation, \cite{pw}
\be
I(gh^{-1})=I(g)+I(h)+{1\over 16\pi}\int d^2\xi Tr[g^{-1}\partial_{\bar z}g h^{-1}\partial_{z}h]
\label{e2}\ee
The associated path integral for $SU(N)_k$ is defined as
\be
Z=\int {\cal D}g~e^{-k~I(g)}
\label{e3}\ee
The (chirally conserved) left and right-moving currents for this theory are given by
\be
J_z=\pp_z g g^{-1}\sp J_{\bar z}=g^{-1}\pp_{\bar z} g\sp \partial_z J_{\bar z}=\pa_{\bar z} J_z=0~.
\label{e4}\ee
We can couple external sources, $A_z,A_{\bar z}$ to them and define their effective action as
\be
Z(A_z,A_{\bar z})=e^{-W(A_z,A_{\bar z})}={1\over Z}\int {\cal D}g~e^{-k~I(g,A_z,A_{\bar z})}
\label{e5}\ee
with
\be
I(g,A_z,A_{\bar z})=I(g)+{1\over 16\pi}Tr\int d^2\xi\left[A_{\bar z}J_z-A_z J_{\bar z}+A_{\bar z}g A_{z} g^{-1}\right]~.
\label{e6}\ee
We may now parameterize without loss of generality the two sources $A_{z,\bar z}$ in terms of two scalar functions, $h,\bar h$,
\be
A_z=\partial_{z}h ~h^{-1}\sp A_{\bar z}=\pp_{\bar z} \bar h ~\bar h^{-1}
\label{e7}\ee
This does not imply, in particular, that the sources have a flat field strength. On the other hand we may write the group elements as non-local functions of the gauge fields
\be
h(z,\bar z)=P~\exp\left[\int_{C_z} A_zdz\right]\sp \bar h(z,\bar z)=P~\exp\left[\int_{C_z} A_{\bar z}d\bar z\right]
\ee
where $C_z$ is a path that links the point $(z,\bar z)$ to a reference fixed point.

Using this parametrization we may rewrite the source action as
\be
I(g,A_z,A_{\bar z})=I(\bar h^{-1}gh)-I(\bar h^{-1} h)+{1\over 16\pi}Tr\int d^2\xi A_{z}A_{\bar z}
\label{e8}\ee
We may now perform the path integral in (\ref{e5}), by changing variables from $g\to \bar h^{-1}gh$ and noting that the path integral measure is invariant under left and right group transformations to obtain
\be
W(A_z,A_{\bar z})=-kI(\bar h^{-1} h)+{k\over 8\pi}Tr\int d^2\xi A_{z}A_{\bar z}
\label{e9}\ee
$$
=-k~I(\bar h^{-1})-k~I(h)=W(A_{\bar z})+W(A_z)
$$
where in the second step we used (\ref{e2}).
This is the final factorized action for the sources. Variation with respect to $A_z,A_{\bar z}$ will give the current correlators.

Note that the source functional $W(A_z,A_{\bar z})$ thus defined is almost gauge invariant under the gauge transformations of the vector sources
\be
A^U_z=UA_zU^{-1}+\pa_zU U^{-1}\sp A^U_{\bar z}=UA_{\bar z}U^{-1}+\pa_{\bar z}U U^{-1}
\label{ee10}\ee
where $U$ is an arbitrary group element depending on $z,\bar z$.
The gauge transformations  correspond in the parametrization of (\ref{e7}) to the transformations
\be
h\to Uh\sp \bar h\to U\bar h
\label{ee11}\ee

The non-gauge invariance of $WA_z,A_{\bar z}$ is a reflection of the two-dimensional
anomaly. The modified functional
\be
\tilde W(A_z,A_{\bar z})=W(A_z,A_{\bar z})-{k\over 8\pi}Tr\int d^2\xi A_{z}A_{\bar z}
\label{ee12}\ee
is gauge invariant as is obvious from (\ref{e9}) and (\ref{ee11}).
Therefore,
\be
W(A^U_z,A^U_{\bar z})-W(A_z,A_{\bar z})={k\over 8\pi}Tr\int d^2\xi
\left[U^{-1}\pa_z U A_{\bar z}+U^{-1}\pa_{\bar z} U A_{z}+U^{-1}\pa_z UU^{-1}\pa_{\bar z} U\right]
\label{ee13}\ee

\section{Holography on $AdS_3$}

There has been a lot of evidence for the holographic correspondence between 2d CFTs and string theories on AdS$_3$ (for a review see \cite{Kraus:2006wn}).

Three-dimensional gravitational theories and their solutions are characterized by a Planck scale
$M_p=1/(16\pi G_3)$ and the associated Planck length, $\ell_p=G_3$.
The solutions are also characterized by their mass $M$ and angular momentum $J$.
The central charge of the corresponding CFT$_2$ is related to the gravity data as
\be
\label{fundaa}
c={3\ell\over 2\ell_p}
~,
\ee
where  $\ell$ is the AdS$_3$ radius.
The general formula for the central charge \eqref{fundaa} was derived in \cite{Brown:1986nw}
using the low-energy gravity description.
Mass and angular momentum are related to the CFT data by
\be
\label{fundab}
M={L_0+\bar L_0\over \ell}\sp J=L_0-\bar L_0~.
\ee
The calculation of the central charge in the gravity theory is expected to be reliable in the semi-classical regime $\ell\gg \ell_p$. Therefore, the result \eqref{fundaa} should be viewed as the leading term in
an expansion in $\ell_p/\ell$. The formulae \eqref{fundab}, however, are expected to be a universal feature of the correspondence.

Assuming that the full string theory dual spacetime is of the form
\beq
\label{fundac}
AdS_3\times \MM_p
~,
\eeq
where $\MM_p$ is a $p$-dimensional compact manifold with volume
$V=\left(\frac{\ell_\MM}{\ell_s}\right)^p$, the three-dimensional Planck length can
be written in terms of the string coupling $g_s$ and the string length $\ell_s$ as
\beq
\label{fundad}
\frac{1}{\ell_p}=\frac{V}{g_s^2 \ell_s}=\frac{1}{g_3^2 \ell_s}
\eeq
where $g_3$ is the three-dimensional string coupling.

Before going further, it is worth recalling the more familiar situation of the $F1$-$NS5$ system
(with $p$ $F1$'s and $k$ $NS5$'s) that involves string theory on $AdS_3\times S^3$. In this case
\beq
\label{fundam}
c=6kp=\frac{3\ell}{2\ell_p}~, ~~ \frac{\ell_s}{\ell_p}=\frac{V}{g_s^2}=\frac{1}{g_3^2}=4p\sqrt{k}
~, ~~ \frac{\ell}{\ell_s}=\sqrt k
~.
\eeq
The volume of the internal manifold $S^3$ and $g_s$ behave as
\beq
\label{fundan}
V\sim \left(\frac{\ell}{\ell_s}\right)^3=k^{3/2}\sp g_s \sim \sqrt{\frac{k}{p}}
~.
\eeq
Defining
\beq
\label{fundap}
N\equiv \sqrt{kp}~~ ({\rm so~that}~c\sim N^2)~~ {\rm and}~~ \lambda\equiv g_s N
\eeq
we obtain
\beq
\label{fundaq}
\lambda=k~~{\rm and}~~ V(\lambda)\sim \lambda^{\frac{3}{2}}
~.
\eeq

If we assume the validity of \eqref{fundaa} for the dual of the $SU(N)_k$ WZW model we
obtain
\be
\label{fundae}
\frac{\ell}{\ell_p}=\frac{2}{3}c\simeq {2N^2\over 3(1+\l)}+{\cal O}(1)
~.
\ee

The next ingredient is the relation of the string coupling to $\lambda$.
To obtain some intuition we will try to establish first in which region we expect a gravitational description.
Recall that the scaling dimensions of representations with a finite number of boxes were given in
(\ref{5}) as
\be
\Delta={\l\over \l+1}\Delta_{R}(\infty)+n+\bar n
\label{e10}\ee
where the integers $n,\bar n$ are the contributions of the current oscillators.
We would like to study the gap in dimensions between the primaries and their descendants. In the limit $\l\to \infty$, the primary dimensions are half integers and therefore there is no adjustable gap separating them from  the excited states. On the other hand, as $\l\to 0$ the primary field dimensions vanish and this creates an adjustable gap. This suggests that the gravity limit will be reliable when $\l\to 0$.
We will adjust $\ell_s$ so that the scaling dimensions in the field theory limit are constant while the excited states' dimensions vary with $\l$. This gives
\be
{\ell\over \ell_s}=\sqrt{1+\l\over \l}
\label{e11}\ee
We may then estimate the three-dimensional string coupling as
\be
g_s^2\sim {\ell_p\over \ell_s}\sim {(1+\l)^{3\over 2}\over N^2\sqrt{\l}}~~\to~~g_sN\sim \left[{(1+\l)^3\over \l}\right]^{1\over 4}~.
\label{e12}\ee
The equation above suggests that the analog of $g_sN$ is never small, both for large and small $\l$.
We have made several assumptions to derive the previous results, including the fact that the string theory dual is a three-dimensional non-critical string theory. This is motivated from the realization of the CFT and previous experience that suggests that extra adjoint matter will generate extra dimensions, whereas fundamental matter is induced by space-filling flavor branes. In that sense, $g_sN$ is a
schematic notation for whatever parameter in the underlying string theory controls the spectrum.

We expect that the bulk string theory has a trivial closed string sector. The reason is two-fold. First, the closed string sector should contain the pure gauge theory states and these are trivial in two dimensions.
Second, as we will see, the open string sector associated with flavor will generate the necessary correlators of the stress tensor and other closed string fields. This does not imply that there is no non-trivial gravitational action for the metric but that there will be no non-trivial fluctuations here and no stringy states.

The non-trivial string sectors are associated with the flavor symmetry $SU(N)_L\times SU(N)_R$. This should be a symmetry that is realized as a bulk gauge symmetry. It will be realized by two sectors of open strings associated to $N$ $D_2$ branes and $N$ $\bar D_2$ as in higher-dimensional realizations of flavor. The gauge fields $L_{\m}$, $R_{\m}$ associated with flavor symmetry will have an action that starts with the CS action as
\be
S_{bulk}={ik\over 8\pi}Tr\int (L dL+{2\over 3}L^3- R dR-{2\over 3}R^3)+\cdots
\label{e13}\ee
where the ellipsis indicates higher derivative terms starting from with the YM action and $L=L^a T^a$
with $Tr[T^aT^b]=\delta^{ab}$. The sign of the coupling constants is implied by the parity invariance of
the CFT. We take $k>0$ without loss of generality.

In Poincar\'e coordinates the AdS$_3$ metric reads
\be
ds^2 =  {\ell^2\over r^2}\left(dt^2+dr^2+dx^2\right) ~.
\label{e14}\ee

The asymptotic expansion near the boundary, $r\to 0$, for solutions of the gauge fields
is\footnote{In appendix \ref{csym} we give a detailed analysis of possible boundary conditions and solutions to the equations of motion for gauge fields in AdS$_3$.} ,
\be
A_i(r,\vec x)=A_i^{(0)}(\vec x)+{{A_i^{(1)}(\vec x)}}r^2 +{\cal O}(r^4)
\label{e15}\ee
where we chose the gauge $A_r=0$.
The equations of motion, including the higher order terms  imply that $A^{(0)}$ is flat
\be
F(A^{(0)})_{ij}=0~.
\label{e16}\ee
They also imply that it is only the CS terms that contribute to boundary terms.
In particular the boundary current is
\be
\Delta S =
{i\over 2\pi}Tr\int d^2x \sqrt{\hat g}J^{i}\delta A_i^{(0)}
\label{e17}\ee
where $\hat g$ is the induced metric on the boundary.

The correct variational principle states that in complex boundary coordinates one of the two
$A_{z,\bar z}$ should be an independent dynamical variable at the boundary.
As described in detail in appendix \ref{csym},  the boundary action that imposes this condition is
\be
S_{\rm boundary}=-{k\over 16\pi}Tr\int_{\partial M} d^2x \sqrt{\hat g}\hat g^{ab}(L_aL_b+R_aR_b)~.
\label{e19}\ee
With this boundary action $L_z$ and $L_{\bar z}$ are canonically conjugate. In particular, one of them can be considered as a source, and the other as the vev.
Which is which depends on the sign of the integer $k$ as seen in (\ref{dd16b}) and (\ref{dd16c}).

Therefore $L^{(0)}_z$ is the source for the anti-holomorphic currents $J_{\bar z}$ of the WZW model, and  $L^{(0)}_{\bar z}$ is proportional to their one-point function,
\be
\langle J_{\bz}\rangle \equiv -2\pi{\delta S\over \delta L^{(0)}_{z}}={k\over 2}L^{(0)}_{\bz}
\label{e18a}\ee
Similarly $R^{(0)}_{\bz}$ is the source for the holomorphic currents $J_{z}$ of the WZW model, and  $R^{(0)}_{z}$ is proportional to their one-point function,
\be
\langle J_{z}\rangle \equiv -2\pi{\delta S\over \delta R^{(0)}_{\bz}}={k\over 2}R^{(0)}_{z}
\label{e18b}\ee

\subsection{The effective action}

We will now establish the effective action for the sources, and match eventually to the one obtained in section \ref{ea} equation (\ref{e9}).
To do this we follow  the analysis in \cite{elitzur}. In this work, it is shown how one can write the gauge fields on-shell, in terms of a two-dimensional flat connection. The end result will be that  the on-shell value of the effective action agrees with (\ref{e9}).

The equations of motion of the CS theory imply that
\be
F_L=F_R=0
\label{e21}\ee
in the bulk whose solution is
\be
L_{\m}=\partial_{\m}h~h^{-1}\sp R_{\m}=\partial_{\m}\bar h~\bar h^{-1}
\label{e22}\ee
with $h,\bar h$ functions of $r$ and the two dimensional coordinates $z,\bar z$.

Evaluating the CS action on the  solution of (\ref{e21}) we obtain
 \be
 S_{bulk}^{\rm on-shell}=-{ik\over 24\pi}\int (L^3-R^3)=-k[\Gamma(h)-\Gamma(\bar h)]=-k[\Gamma(h)+\Gamma(\bar h^{-1})]
\label{e23}\ee
Adding this to the boundary action (\ref{e19}) where $h,\bar h$ are evaluated at the boundary, we obtain perfect agreement with the effective action obtained from the CFT (\ref{e9}).

We may also compute the boundary stress tensor.
The bulk action does not contribute as the CS term is metric independent, and the YM term gives vanishing boundary contributions as explained in appendix \ref{csym}.
The only contribution comes from the counterterm,
\be
T^b_{ij}\equiv -{4\pi\over \sqrt{\hat g}}{\delta S_{on-shell}\over \delta \hat g^{ij}}={|k|\over 2}Tr\left[L_{i}L_{j}-{1\over 2}\hat g_{ij}\hat g^{ab}L_{a}L_{b}+(L\leftrightarrow R)\right]
\label{dd36}\ee
which is traceless
\be
T_{zz}={k\over 2}Tr[L^{(0)}_zL^{(0)}_z+R^{(0)}_z R^{(0)}_z]\sp T_{\bz\bz}={k\over 2}Tr[L^{(0)}_{\bz}L^{(0)}_{\bz}+R^{(0)}_{\bz} R^{(0)}_{\bz}]\sp T_{z\bz}=0
\label{dd37}\ee
Using (\ref{e18a}), (\ref{e18b}) and setting the current sources $L^{(0)}_z, R^{(0)}_{\bz}$ to zero we obtain
\be
T_{zz}={2\over k}Tr[J_{z}J_{z}]\sp T_{\bz \bz}={2\over k}Tr[J_{\bz}J_{\bz}]
\label{ee20}\ee
which is the (unrenormalized) Sugawara stress tensors.

As mentioned before, we should think of $L^{(0)}_{\bz}$ as the current induced by the source $L^{(0)}_z$.
Due to (\ref{e21}) they satisfy
\be
\pa_zL^{(0)}_{\bz}-\pa_{\bz}L^{(0)}_z+[L^{(0)}_z,L^{(0)}_{\bz}]=0
\label{ee21}\ee
with solution
\be
L^{(0)}_z=h^{-1}\pa_z h\sp L^{(0)}_{\bz}=h^{-1}\pa_{\bz} h={2\over k}\langle J_{\bar z}\rangle
\label{ee22}\ee
We may write
\be
h=P~\exp\left[\int_{C_{\infty}}^{(z,\bz)}L^{(0)}_zdz\right]
\label{ee23}\ee
where the path starts from a reference point $C_{\infty}$ and ends up at $(z,\bz)$.
We may then compute
\be
L^{(0)}_{\bz}=h^{-1}\pa_{\bz}h=P~\exp\left[-\int_{C_{\infty}}^{(z,\bz)}L^{(0)}_zdz\right]
\int_{C_{\infty}}^{(z,\bz)}\pa_{\bz}L^{(0)}_zdz~~P~\exp\left[\int_{C_{\infty}}^{(z,\bz)}L^{(0)}_zdz\right]
\ee
$$
=\int_{C_{\infty}}^{(z,\bz)}\pa_{\bz}L^{(0)}_zdz+{\cal O}\left(\left[L^{(0)}_z\right]^2\right)
$$

Then the relation between source and current becomes to order ${\cal O}\left(\left[L^{(0)}_z\right]^2\right)$
\be
\langle J_{\bar z}\rangle ={k\over 2}L_{\bz}={k\over 2}h^{-1}\pa_{\bz} h={k\over 2}\int_{C_{\infty}}^{(z,\bz)}\pa_{\bz}L^{(0)}_z~dz
\label{ee24}\ee
from where we obtain the response function
\be
{\delta J^a_{\bz}(z)\over \delta L^{(0)b}_z(w)}={\delta \over \delta L^{(0)b}_z(w)}{k\over 2}\int_{C_{\infty}}^{(z,\bz)}\pa_{\bz}L^{(0)a}_zdz={k\over 2}
\delta^{ab}\int_{C_{\infty}}^{(z,\bz)}\pa_{\bz}\delta^{(2)}(z-w)dz={k\delta^{ab}\over 4\pi}\int_{C_{\infty}}^{(z,\bz)}\pa_{\bz}\pa_z{1\over \bz-\bar w}dz
\label{ee25}\ee
$$
={k\delta^{ab}\over 4\pi}\int_{C_{\infty}}^{(z,\bz)}\pa_z\pa_{\bz}{1\over \bz-\bar w}dz=-{k\delta^{ab}\over 4\pi}{1\over (\bz-\bar w)^2}
$$
which is the proper two point function of the anti-holomorphic currents.

Equation (\ref{ee21}) can be solved iteratively in powers of the source $L^{(0)}_z$ by expanding
\be
L^{(0)}_{\bz}\equiv \sum_{n=1}^{\infty} L^n_{\bz}\sp \pa_z L^1_{\bz}=\pa_{\bz}L^{(0)}_z\sp \pa_{z}L^n_{\bz}=[L^{n-1}_{\bz},L^{(0)}_z]\sp n\geq 2
\label{ee25}\ee
with solution
\be
L^1_{\bz}=\int _{C_{\infty}}^{(z,\bz)}\pa_{\bz}L^{(0)}_zdz\sp L_{\bz}^n=\int _{C_{\infty}}^{(z,\bz)}[L^{n-1}_{\bz},L^{(0)}_z]dz
\ee
From $L^n_{\bz}$ we can construct the n-point function of anti-holomorphic currents as
\be
\langle \prod_{i=1}^{n}J^{a_i}_{\bz}(w_i)\rangle= \left(\prod_{i=1}^{n} {\delta \over \delta L^{(0)a_i}_{\bz}}\right)~L^{n-1}_{\bz}
\ee

\subsection{The bulk scalar}

{From} the non-trivial ground states of the WZW theory associated to primaries, only one can be considered as the generating operator dual to a complex bulk scalar $T_{ij}$, that transforms in the bi-fundamental under the bulk gauge group $SU(N)_L\times SU(N)_R$. The reason is that all other primary ground states can be considered as composites (multi-particle states) of the fundamental scalar $g$ under OPE, since they arise as appropriately regularized algebraic functions (products) of the fundamental operator.

Its mass is given by the standard formula that connects it to the scaling dimension of the dual operator
\be
h=\Delta+\bar \Delta={\l\over \l+1}+{\cal O}\left({1\over N^2}\right)\sp m^2\ell^2=-{\l(\l+2)\over (\l+1)^2}
~.
\label{e24}\ee
The situation is similar to that of tachyon condensation in $D-\bar D$ systems realizing flavor in holography, \cite{tachyon}.
The simplest two-derivative quadratic action compatible with the symmetries is
\be
S_{T}={1\over 2}Tr\int \sqrt{g}\left(g^{\m\n}D_{\m}TD_{\n}T^{\dagger}+m^2TT^{\dagger}\right)+{h\over \ell}Tr\int_{\partial M} d^2x \sqrt{\hat g}~ TT^{\dagger}
\label{e25}\ee
where the last term is the standard quadratic boundary counterterm and
\be
D_{\m}T=\partial_{\mu}T+iL_{\m}T-iTR_{\m}\sp D_{\m}T^{\dagger}=\partial_{\mu}T^{\dagger}+iR_{\m}T^{\dagger}-iT^{\dagger}L_{\m}~.
\label{e26}\ee
The vev associated with the CFT vacuum is $T=0$, which is the only choice that keeps the chiral symmetry unbroken.
This action will have higher order corrections in $T$ in order to generate the multipoint correlation functions of the WZW theory.
It is plausible that the full non-linear theory is described by a DBI-like action along the lines of \cite{sen}, \cite{garousi}.

Turning on a source for $T$ should correspond to perturbations of the WZW theory by $g$.
In the strong coupling limit this corresponds to turning on a mass matrix for the fermions.
Therefore, the theory is expected to flow to $SU(N-r)_k$  theory where $r$ is the rank of the mass-matrix.
This amounts to a reduction of the values of the 't Hooft coupling $\l$, therefore flowing towards weak coupling.
The flow is also visible in the bulk theory, (\ref{e13}), (\ref{e25}).

\subsection{Multiparticle states  and entropy}

We have seen in section \ref{su(N)k} that ground states that are multiparticle states can have large multiplicities. Their scaling dimensions are of order ${\cal O}(N^2)$ and therefore may be expected to correspond to macroscopic solutions in the bulk theory. Since however the group degrees of freedom are unconfined (and therefore observable), such states would not correspond to highly entropic states as they are distinguished by their $SU(N)^2$ quantum numbers. It is therefore expected, but not
completely clear, that they should correspond to regular horizonless semiclassical bulk solutions.

The associated mass for the ``half-box" states in (\ref{9}) is given by
\be
{M\over M_p}={\Delta+\bar \Delta \over {N^2\over 24\pi (1+\l)}}\simeq {6\pi}{\l+1\over \l}
\ee
where we have used eqs.\ (\ref{fundaa}), (\ref{fundab}).
All other primary masses are suppressed by 1/$N$ or more.
Observe that in the weak coupling limit, $\l\to 0$, the mass becomes much larger than the
Planck scale, ${M\over M_p}\sim {1\over \l}$. Hence, these are truly macroscopic configurations.
We recall that this is also the limit where the zero mode theory (gravity+CS+scalar) is expected
to give a reliable description of the physics.

On the other hand, for $\l\to\infty$, the mass of these states is bounded, ${M\over M_p}\sim {6\pi}$
by the Planck scale. Therefore, in this limit the ground states remain Planckian and do not
generate macroscopic states.

The multiplicity of the ``half-box'' state was calculated in (\ref{10}). We can compare it with the Cardy entropy for a state with the same scaling dimension as the conformal dimensions $\Delta=\bar \Delta={N^2\over 8\l}$ in (\ref{7})
\be
S_{Cardy}=2\cdot 2\pi\sqrt{c\Delta\over 6}\simeq {\pi\over \sqrt{3}}{N^2\over \l}+\cdots~.
\label{13}\ee
We observe that at strong coupling the dimension multiplicity in (\ref{11}) is subleading to (\ref{13})
as ${\pi\over \sqrt{3}}>2\log 2$. Such states have subdominant entropy, and never dominate the
Cardy entropy. On the other hand, at weak coupling the Cardy entropy is subdominant. This suggests
that in this intermediate energy regime it is this class of states that dominate the partition function. Of
course for asymptotically high energies it will be the Cardy entropy that will dominate.

We conclude with the following picture. The theory possesses ground states of different scaling
dimensions and degeneracy. Some of them, like the half-box states, are highly massive and can
dominate the partition function, but do not give rise to highly entropic states in the bulk. There are
also massive states with high entropy at the Regge trajectories of each ground state. These are
expected to correspond to bulk black holes. The large degeneracy of some ground states, like
the half-box states, implies the existence of a correspondingly large class of ultra-massive black
holes in the bulk.

\section*{Acknowledgements}\label{ACKNOWL}

We would like to thank O.\ Aharony, M.\ Douglas, R.\ Gopakumar, P.\ Kraus, K.\ Papadodimas and C. Sonnenschein for discussions and correspondence.
 E.K.\ thanks the ESI institute in Vienna and the GGI institute in Firenze for hospitality during the completion of this work.

This work was partially supported by the European Union grants FP7-REGPOT-2008-1-CreteHEPCosmo-228644 and PERG07-GA-2010-268246.

\newpage

\renewcommand{\theequation}{\thesection.\arabic{equation}}
\addcontentsline{toc}{section}{Appendices}

\section*{Appendices}

\begin{appendix}

\section{Casimirs and dimensions for the $SU(N)$ algebra\label{casimir} }

In this appendix we will tabulate some useful results on $SU(N)$ representations namely
dimensions, the Dynkin index and the quadratic Casimir that enters in conformal dimensions
of primary fields.

We will denote the dimension of the representation $R$ by $D(R)$. The irreducible representation
$R$ is completely specified with its Young tableau. Conjugation by an $\epsilon$-tensor acts on any
column of length $s\leq N$ of the Young tableau replacing it with a column of length $N-s$.
We use the convention to have the minimal number of boxes for an $SU(n)$ representation.  For example, if we discuss the $\overline{\Yboxdim8pt\yng(1)}$ that can be described also by $N-1$ antisymmetrized boxes we substitute instead the fundamental with one box.
Therefore, a representation represented by a generic Young tableau has $m_1$ boxes in the first column, $m_2$ boxes in the second column etc, with $m_1\geq m_2\geq m_3\cdots \geq m_n$.
We will denote this representation as $(m_1,m_2,\cdots, m_n)$.

The affine cutoff implies that only integrable representations can be present as primary states.
For $SU(N)_k$, all representations with more than $k$ columns are not integrable. Therefore
$n\leq k$ above.

In a tensor product of irreducible representations $R_1\otimes R_2=\sum_k R_k$
we have the following relation for their  dimensions
\be
D(R_1)D(R_2)=\sum_{k} D(R_k)
\label{b1}\ee
from which the well known formulae for dimensions of an arbitrary Young tableau can be calculated.

The Dynkin index $S_2(R)$ of a representation $R$ is defined as, \cite{patera}
\be
Tr[T^a_R T^b_R]=S_2(R)~\delta ^{ab}
\label{b2}\ee
where $T^a_R$ are the Lie algebra generators of the representation $R$.
We will normalize them here  so that $S_2(\Yboxdim6pt\yng(1)\hskip 2pt)={1\over 2}$.
Again for a tensor product we have the following relations that allow the calculation of all Dynking indices
\be
D(R_1)S_2(R_2)+D(R_2)S_2(R_1)=\sum_k S_2(R_k)~.
\label{b3}\ee

Finally the quadratic Casimir is defined as
\be
\sum_a(T^a_RT^a_R)_{ij}=C_2(R)\delta_{ij}
\label{b4}\ee
and is related to the Dynking index by
by
\be
(N^2-1)S_2(R)=D(R)C_2(R)
\label{b5}\ee
that follows from (\ref{b2}) and (\ref{b4}).

The general formula for the Casimir for an arbitrary representation $(m_1,m_2,\cdots, m_n)$ is
\be
C_2(m_1,m_2,\cdots)={(\sum_{i=1}^nm_i)N^2-((\sum_{i=1}^n m_i^2)
-\sum_{i=1}^n (2i-1)m_i)N-(\sum_{i=1}^nm_i)^2\over 2 N}.
\label{b6}\ee
The quadratic Casimir is related to the conformal dimensions of primary fields as
\be
\Delta_R=\bar \Delta_R={C_2(R)\over k+N}~.
\label{b7} \ee
The large-$N$ limit of dimensions is obtained from the large-$N$ limit of the Casimir
\be
\lim_{N\to \infty} C_2(R)=N\Delta_R(\infty)+{\cal O}(1)\sp \lim \Delta_R
={\l \over 1+\lambda}\Delta_R(\infty)
\label{b8}\ee
with the 't Hooft coupling defined in (\ref{a19}).
We obtain
\be
\Delta_R(\infty)={1\over 2}(\sum_{i}m_i)
\label{b9}\ee
for $m_i\sim {\cal O}(1)$.

The dimension, Dynkin index, Casimir and $\Delta_R(\infty)$ for some common representations are tabulated in table \ref{table1}.

\begin{center}
\begin{tabular}{|c|c|c|c|c|} \hline
Representation & dimension & Dynkin Index $S_2$ & Casimir $C_2$& $\Delta_R(\infty)$\\
\hline \hline

$\Yboxdim8pt\yng(1)$ & $N$ & ${1\over 2}$ & ${N^2-1\over 2N}$&${1\over 2}$\\

$\Yboxdim8pt\yng(2)$   &  ${N(N+1)\over 2}$ & ${N+2\over 2}$ & ${(N-1)(N+2)\over N}$&1\\

$\Yboxdim8pt\yng(1,1)$   &  ${N(N-1)\over 2}$ & ${N-2\over 2}$ &${(N+1)(N-2)\over N}$&1\\

Adjoint &  $N^2-1$& $N$&$N$&1\\

$\Yboxdim8pt\yng(3)$& ${N(N+1)(N+2)\over 6}$& ${(N+2)(N+3)\over 4}$&${3(N-1)(N+3)\over 2N}$ &${3\over 2}$\\

$\Yboxdim8pt\yng(2,1)$& ${N(N^2-1)\over 3}$& ${N^2-3\over 2}$&${3(N^2-3)\over 2N}$& ${3\over 2}$\\

$\Yboxdim8pt\yng(1,1,1)$& ${N(N-1)(N-2)\over 6}$& ${(N-2)(N-3)\over 4}$&${3(N+1)(N-3)\over 2N}$  &${3\over 2}$\\

$\Yboxdim8pt\yng(4)$& ${N(N+1)(N+2)(N+3)\over 24}$& ${(N+2)(N+3)(N+4)\over 12}$&${2(N-1)(N+4)\over N}$ &2\\

$\Yboxdim8pt\yng(1,1,1,1)$& ${N(N-1)(N-2)(N-3)\over 24}$& ${(N-2)(N-3)(N-4)\over 12}$&${2(N+1)(N-4)\over N}$ &2\\

$\Yboxdim8pt\yng(2,2)$& ${N^2(N^2-1)\over 12}$& ${(N+2)(N+3)(N+4)\over 12}$&${2(N^2-4)\over N}$&2 \\

$\Yboxdim8pt\yng(3,1)$& ${N(N-1)(N+1)(N+2)\over 8}$&  ${(N+2)(N^2+N-4)\over 4}$&${2(N^2+N-4)\over N}$ &2\\

$\Yboxdim8pt\yng(2,1,1)$& ${N(N+1)(N-1)(N-2)\over 8}$&  ${(N-2)(N^2-N-4)\over 4}$&${2(N^2-N-4)\over N}$&2 \\

m-symmetric & $\left(N+m-1\atop m\right)$& ${1\over 2} \left(N+m\atop m-1\right)$  &${m(N-1)(N+m)\over 2N}$&  ${m\over 2}$ \\

m-antisymmetric & $\left(N\atop m\right)$& ${1\over 2} \left(N-2\atop m-1\right)$  &${m(N-m)(N+1)\over 2N}$ & ${Min[m,N-m]\over 2}$\\
\hline
\end{tabular}
\label{table1}
\end{center}

We will now consider representations with large dimensions and large multiplicities.
The maximum is  when
\be
m_i={N\over 2}+n_i\sp n_i\sim {\cal O}(1)
\label{b10}\ee
for which we compute
\be
C_2={Nk(N+k)\over 8}-{\sum_{i=1}^k n_i^2+\sum_{i=1}^k (k+1-2i)n_i+{1\over N}(\sum_{i=1}^k n_i)^2\over 2}
\label{b10a}\ee
and the scaling dimension
\be
\Delta={C_2\over N+k}={Nk\over 8}-{\sum_{i=1}^k n_i^2+\sum_{i=1}^k (k+1-2i)n_i+{1\over N}(\sum_{i=1}^k n_i)^2\over 2(N+k)}
=
\label{b11}\ee
$$={N^2\over 8\lambda}-{\lambda\over 2(\lambda+1)}\left[{1\over N}\sum_{i=1}^k n_i^2+
{1\over N}\sum_{i=1}^k (k+1-2i)n_i+{1\over N^2}(\sum_{i=1}^k n_i)^2\right]~.
$$
The dimension of the associated $SU(N)$ representation when $n_i=0$ for all $i$ is
\be
D={\prod_{i=1}^{N/2}{(N+k-i)!\over (N-i)!}\over \prod_{i=1}^{k}{\left({N\over 2}+i-1\right)!\over (i-1)!}}
\label{b12}\ee

The logarithm of this dimension is
\be
\log D=\sum_{i=1}^{N/2}\left[\log (N+k-i)!-\log (N-i)!\right]- \sum_{i=1}^{k}\left[\log\left({N\over 2}+i-1\right)!-\log  (i-1)!\right]~.
\label{b13}\ee
Using Stirling's formula
\be
\log(n!)=n\log n-n+{1\over 2}\log (2\pi n)+{\cal O}(n^{-1})
\label{b14}\ee
we obtain
\be
\log D=\sum_{i=1}^{N/2}\left[\left(N-i+{1\over 2}\right)  \log{(N+k-i)\over (N-i)} + k\log(N+k-i)\right]-
\label{b15}\ee
$$-
\sum_{i=1}^{k}\left[{N\over 2}\log\left({N\over 2}+i-1\right)+\left(i-{1\over 2}\right)\log{\left({N\over 2}+i-1\right)\over (i-1)}
\right]~.
$$
We may now use the summation formula
\be
\sum_{i=1}^N f(i)=\int_1^{N+1}f(x)dx+{1\over 2}(f(N+1)-f(1))+{1\over 12}(f'(N+1)-f'(1))-{1\over 72}(f''(N+1)-f''(1))+{\cal O}(f''')
\label{b16}\ee
to obtain
\be
\log D=\left[4(\l+1)\log 2-\l^2\log \l+2(\l+1)^2\log(\l+1)-(\l+2)^2\log(\l+2)\right]{N^2\over 4\l^2}+
\label{b17}\ee
$$
-{N\over 2}\log N+\left[2\log(\l+2)-4(\l+1)\log (\l+1)+2(\l-1)\log \l+2\l \log 2\right]{N\over 4\l}
-
$$
$$
-{1\over 12}\log{\l+2\over \l}+{\cal O}(N^{-1})~.
$$

\section{Free fermions,  $O(2N)_1$,  $U(N)_1$, and $SU(N)_k$ \label{free}}

Consider $N$ complex free left-moving fermions $\psi^i,\bar\psi^i$, $i=1,...,N$, equivalent to $2N$
real (Majorana-Weyl) ones. They realize the $O(2N)_1$ current algebra, \cite{wzw}. This is equivalent as a CFT to $U(N)_1\sim U(1)\times SU(N)_1$.

The $O(2N)_1$ theory contains the unit affine representation, the vector (V) representation
(of dimension $2N$), the spinor (S) of dimension $2^{N-1}$,
and the conjugate spinor (C) with dimension also $2^{N-1}$.

The left-right symmetric character-valued partition function is
\be
Z_{O(2N)_1}(v,\bar v)={1\over 2}\sum_{a,b=0}^1\;\;\prod_{i=1}^N{|\vartheta[^a_b](v_i)|^2\over |\eta|^2}=|\chi_{0}|^2+|\chi_{V}|^2+|\chi_{S}|^2+|\chi_{C}|^2
\label{c1}\ee
with  $O(2N)_1$ characters
\be
\chi_{0}(\vec v)={1\over 2}\left[{\prod_{i=1}^{N}\vartheta_{3}(v_i)\over \eta^N}+{\prod_{i=1}^N\vartheta_{4}(v_i)\over \eta^N}\right]
=q^{-N/24}[1+O(q)]~,
\label{c2}\ee
\be
\chi_{V}(\vec v)={1\over 2}\left[{\prod_{i=1}^{N}\vartheta_{3}(v_i)\over \eta^N}-{\prod_{i=1}^N\vartheta_{4}(v_i)\over \eta^N}\right]
=2N\;q^{-{N\over 24}+{1\over 2}}[1+O(q)]~,
\label{c3}\ee
\be
\chi_{S}(\vec v)={1\over 2}\left[{\prod_{i=1}^{N}\vartheta_{2}(v_i)\over \eta^N}+{\prod_{i=1}^N\vartheta_{1}(v_i)\over \eta^N}\right]=2^{N-1}
q^{-{N\over 24}+{N\over 8}}[1+O(q)]~,
\label{c4}\ee
\be
\chi_{C}(\vec v)={1\over 2}\left[{\prod_{i=1}^{N}\vartheta_{2}(v_i)\over \eta^N}-{\prod_{i=1}^N\vartheta_{1}(v_i)\over \eta^N}\right]=2^{N-1}
q^{-{N\over 24}+{N\over 8}}[1+O(q)]~.
\label{c5}\ee

We will write the theory in terms of the $U(1)$ and $SU(N)_1$ degrees of freedom.
The representations that descend from the unit and vector of $O(2N)$ are generated by the operators
$\psi^i$, $\psi^i\psi^j$, $\ldots$ ,$\prod_{k=1}^{N}\psi^{i_{k}}$, and their complex conjugates which correspond to the various antisymmetric representations of $SU(N)$. These are the only integrable representations at level one.

The $U(1)\times SU(N)_1$ decomposition at the level of conformal dimensions becomes
\be
\Delta ={Q^2\over 2}+\Delta _{R}
\label{c6}\ee
where $Q$ is the appropriately normalized $U(1)$ charge (defined when the associated current has central term equal to one),
and $\Delta_{R}$ is the conformal weight of the $SU(N)_{1}$ reps.
$$\Delta_{R}={C_{R}\over N+1}$$

For the $j$-index antisymmetric representation of $SU(N)$, $C_{R}=(N+1)j(N-j)/2N$ so that
\be
\Delta_j={j(N-j)\over 2N}
\label{c7}\ee
Its (properly normalized) $U(1)$ charge is $\pm j/\sqrt{N}$.
Thus, summing up the two contributions in eq.\ \eqref{c6} we obtain conformal weight $j/2$
which is in agreement with the interpretation above.

Consider now the two spinor representations.
Their $U(1)$ charges can be easily figured out by bosonizing pairwise the complex fermions
\be
\psi^{i}=e^{i\phi_{i}}\;\;\;,\;\;\;\bar \psi^{i}=e^{-i\phi_{i}}~.
\label{c8}\ee
In this basis, the $U(1)$ current is
\be
J={i\over \sqrt{N}}\sum_{k=1}^{N}\partial \phi_{k}~.
\label{c9}\ee
The spinor and conjugate spinor are generated by the following vertex operators
\be
V_{C,S}=\prod_{k=1}^{N}exp\left[{i\over 2}\epsilon_{k}\phi_{k}\right]
\label{c10}\ee
with $\epsilon_{k}=\pm 1$.
The spinor corresponds to $\prod_{K=1}^{N}\epsilon_{k}=1$
and the conjugate spinor to $\prod_{K=1}^{N}\epsilon_{k}=-1$.
The $U(1)$ charge is $Q=\sum_{k=1}^{N}\epsilon_{k}/2\sqrt{N}$.
It is not difficult to see that the spectrum of $U(1)$ charges coming from C and S
is given by
\be
Q_{k}={N-2k\over 2\sqrt{N}}\;\;\;,\;\;\;k=0,1,2,...,N
\label{c11}\ee
where $k$ even corresponds to one spinor and $k$ odd to the other.
C,S decompose under $U(1)\times SU(N)$ to the antisymmetric reps. This can be confirmed by
the conformal weights. When the $U(1)$ charge is $Q_{k}$, the accompanying $SU(N)$
representation is the $k$-index antisymmetric representation,
\be
{Q_{k}^2\over 2}+{k(N-k)\over 2N}={N\over 8}
\label{c12}\ee
which is the correct conformal weight of the spinors.

The quantum of $U(1)$ charge is $1\over \sqrt{N}$ in the unit-vector sector and
${1\over 2\sqrt{N}}$ in the spinor sector. The number of states in the Dirac spinor is
$2^N=2\cdot 2^{N-1}$. This is the number of all antisymmetric representation states.

\subsection{The fermionic current algebra}

We define the $U(N)$ currents $J^{ij}$ in terms of the fermions that satisfy
\be
\psi^i(z)\psi^j(w)\sim {\it finite} \sp  \bar\psi^i(z)\bar\psi^j(w)\sim {\it finite}\sp
\psi^i(z)\bar \psi^j(w)\sim {\delta_{ij}\over z-w}+ {\it finite}
\label{c13}\ee
\be
J^{ij}=i:\psi^i\bar \psi^j:\sp J^{ij}(z)J^{kl}(w)
={\delta^{il}\delta^{jk}\over (z-w)^2}+i{f^{ij,kl}}_{mn} J^{mn}(w)+{\it finite}
\label{c14}\ee
\be
{f^{ij,kl}}_{mn}=-\delta^{il}\delta^{mk}\delta^{jn}+\delta^{jk}\delta^{mi}\delta^{nl}~.
\label{c15}\ee
The properly normalized overall $U(1)$ current is
\be
J={1\over \sqrt{N}}\sum_iJ^{ii}\sp J(z)J(w)={1\over (z-w)^2}+{\it finite}~.
\label{c16}\ee
The currents are uncharged under the zero mode $J_0$.

Consider now the antisymmetric operator
$O^{i_1,\cdots,i_{m}}=:\psi^{i_1}\psi^{i_2}\cdots\psi^{i_m}$ with
$U(1)$ charge $Q={n\over \sqrt{N}}$
\be
J^{ij}(z)O^{i_1,\cdots,i_{m}}(w)=\sum_{n=1}^m{(-1)^{n+1}\delta^{ji_n}\over z-w}O^{i_1,\cdots,i,\cdots,i_{m}}+{\it finite}~.
\label{c17}\ee
This indicates that the operators $O$ are affine primaries, that transform as
the $m$-index antisymmetric of $SU(N)$.

Of interest is the $m=N$ operator $O=\prod_{i=1}^N \psi^i$ that satisfies
\be
J^{ij}(z)O(w)={\it finite}, ~~~i\not= j~.
\label{c17a}\ee
Indeed, it can be seen that this operator has charge $Q=\sqrt{N}$ and therefore dresses the
trivial $SU(N)$ representation. If one bosonizes the U(1) current in (\ref{c16}) as
\be
J=i\partial\phi
\ee
 then this state is a pure vertex operator
 \be
 O(z)=:e^{i\sqrt{N}\phi}:
 \ee

\subsection{$SU(N)_k$}

We now consider $k$ copies of $N$ complex fermions,  $\psi^i_a$,  $\bar\psi^i_a$,
$a=1,2,\cdots,k$, $i=1,2,\cdots,N$.
\be
\psi^i_a(z)\psi^j_b(w)\sim {\it finite} \sp  \bar\psi^i_a(z)\bar\psi^j_b(w)\sim {\it finite}\sp
\psi^i_a(z)\bar \psi^j_b(w)\sim {\delta_{ij}\delta^{ab}\over z-w}+{\it finite}~.
\label{c18}\ee
They realize the tensor product CFT $U(N)_1^k$.

We may construct the  $U(N)_k$ currents as
\be
J^{ij}=\sum_{a=1}^k :\psi^i_a\bar\psi^j_a:
\sp J^{ij}(z)J^{kl}(w)=k{\delta^{il}\delta^{jk}\over (z-w)^2}+i{f^{ij,kl}}_{mn} J^{mn}(w)+{\it finite}
\label{c19}\ee
as well as $U(k)_N$ currents
\be
J^{ab}=\sum_{i=1}^N :\psi^i_a\bar\psi^i_b:
\sp J^{ab}(z)J^{cd}(w)=N{\delta^{ad}\delta^{bc}\over (z-w)^2}+i{f^{ab,cd}}_{ef} J^{ef}(w)+{\it finite}~.
\label{c20}\ee
The two groups are not independent as they share the same overall $U(1)$
\be
J={1\over \sqrt{kN}}\sum_{i=1}^N\sum_{a=1}^k :\psi^i_a\bar\psi^i_a~.
\label{c21}\ee
Moreover, the two current algebras are not commuting. A general commutator gives other
currents of the maximal $O(2kN)_1$ current algebra.

The global subalgebra of the $U(k)_N$ algebra acts non-trivially on the $(U(N)_1)^k/U(N)_k$ coset.
To leading order it will be a symmetry of $U(N)_k$.

In this respect, the present theory is given by $O(2kN)_1= U(kN)_1\simeq (U(N)_1)^k$ and
\be
O(2kN)_1=SU(N)_k\otimes SU(k)_N\otimes U(1)\sp SU(N)_k={O(2kN)_1\over SU(k)_N\otimes U(1)}~.
\label{c22}\ee

In the limit $N\to\infty$, with $k$ fixed, the coset  $SU(k)_N$ has central charge of order
${\cal O}(1)$. Hence, to leading order
\be
SU(N)_k\simeq O(2kN)_1~.
\label{c23}\ee
The same is true in the 't Hooft limit with $\l\gg 1$.
Therefore, in these limits the WZW model reduces to a theory of $kN$ free complex fermions.

Moreover, (\ref{c22}) can be translated to the statement that the conformal $U(k)$ gauge theory of
$N$ flavors of massless fermions is equivalent to the $SU(N)_k$ WZW model.
In this respect, the 't Hooft limit $k\to\infty$, $N\to\infty$ with $N/k=\l$ fixed can be interpreted as a Veneziano limit with $\lambda$ being the ratio of flavors to colors.

(\ref{c22}) can also be written in the alternative form:
\be
U(N)_k={O(2kN)_1\over SU(k)_N}
\label{c29}\ee
which indicates the full flavor symmetry of an $SU(N)_k$ gauge theory including the $U(1)_{L}\times U(1)_R$ flavor symmetries.
Note that unlike four dimensions the $U(1)_A$ is not anomalous in two dimensions as the anomaly is quadratic in the group generators and it vanishes because of the tracelessness of the generators of the color group.

Several of the above issues were discussed in early papers on the realization of chiral symmetry in two dimensions, \cite{2d,cobi}.

\section{Analysis of a four-point function in $SU(N)_k$\label{4p}}

In this appendix we present the details of the analysis of the $SU(N)_k$ four-point function of
the fundamental $g_{a,b}(z,\bar z)$ and its conjugate $g^{-1}_{b,a}(z,\bar z)$ following \cite{kz}.

The result is
 \be
 \langle g_{a_1,b_1}(z_1,\bar z_1)g^{-1}_{b_2,a_2}(z_2,\bar z_2)g_{a_3,b_3}
 (z_3,\bar z_3)g^{-1}_{b_4,a_4}(z_4,\bar z_4)\rangle
 =|z_{14}z_{23}|^{-4\Delta_{\Yboxdim4pt\yng(1)}} ~G({x,\bar x})
 =
\label{dd1}\ee
 $$
 =\langle g_{a_1,b_1}(\infty)g^{-1}_{b_2,a_2}(1)g_{a_3,b_3}
 (x,\bar x)g^{-1}_{b_4,a_4}(0)\rangle
$$
where $z_{ij}\equiv z_i-z_j$, $x$ is the standard cross-ratio
\be
 x={z_{12}z_{34}\over z_{14}z_{32}}
\label{d2} \ee
 and bars stand for complex conjugation.

The function $G$ can be decomposed into group channels as
\be
G(x,\bar x)
=\sum_{A,B=1}^2 I^A \bar I^B~G_{AB}(x,\bar x)
\label{d3}\ee
with
\be
I^1=\delta_{a_1,a_2}\delta_{a_3,a_4}\sp   \bar I^1=\delta_{b_1,b_2}\delta_{b_3,b_4}
\sp I^2=\delta_{a_1,a_4}\delta_{a_2,a_3}\sp   \bar I^2=\delta_{b_2,b_4}\delta_{b_1,b_3}
\label{d4} \ee
and conformal block channels as
\be
G_{AB}(x,\bar x)={\cal F}^{(1)}_A(x){\cal F}^{(1)}_B(\bar x)+h{\cal F}^{(2)}_A(x){\cal F}^{(2)}_B(\bar x)
~.
\label{d5} \ee
The conformal blocks have been calculated by solving the Knizhnik-Zamolodchikov equations, \cite{kz} and are given by
\be
{\cal F}^{(1)}_1(x)=x^{-2\Delta_{\Yboxdim4pt\yng(1)}}(1-x)^{\Delta_A-2\Delta_{\Yboxdim4pt\yng(1)}}
~F\left(-{1\over 2\kappa},{1\over 2\kappa};1+{N\over 2\kappa},x\right)~,
\label{d6} \ee
\be
{\cal F}^{(1)}_2(x)=-{x^{1-2\Delta_{\Yboxdim4pt\yng(1)}}(1-x)^{\Delta_A-2\Delta_{\Yboxdim4pt\yng(1)}}
\over 2\kappa+N}
~F\left(1-{1\over 2\kappa},1+{1\over 2\kappa};2+{N\over 2\kappa},x\right)~,
\label{d7}\ee
\be
{\cal F}^{(2)}_1(x)=
x^{\Delta_A-2\Delta_{\Yboxdim4pt\yng(1)}}(1-x)^{\Delta_A-2\Delta_{\Yboxdim4pt\yng(1)}}
 ~F\left(-{N-1\over 2\kappa},-{N+1\over 2\kappa};1-{N\over 2\kappa},x\right)~,
\label{d8}\ee
\be
{\cal F}^{(2)}_2(x)
=-Nx^{\Delta_A-2\Delta_{\Yboxdim4pt\yng(1)}}(1-x)^{\Delta_A-2\Delta_{\Yboxdim4pt\yng(1)}}
~F\left(-{N-1\over 2\kappa},-{N+1\over 2\kappa};-{N\over 2\kappa},x\right)~,
\label{d9}\ee
where $F$ is the hypergeometric function,
\be
\Delta_{\Yboxdim4pt\yng(1)}={N^2-1\over 2N(N+k)}\sp \Delta_A={N\over N+k}\sp 2\kappa=-N-k
\label{d10}\ee
are the conformal dimensions of the fundamental and adjoint affine primaries
and the $(\Yboxdim4pt\yng(1),\Yboxdim4pt\yng(1), {\rm Adjoint})$ OPE coefficient $h$ is given by
\be
h={1\over N^2}{\Gamma\left[{N-1\over N+k}\right]\Gamma
\left[{N+1\over N+k}\right]\Gamma^2\left[{k\over N+k}\right]
\over \Gamma\left[{k+1\over N+k}\right] \Gamma\left[{k-1\over N+k}\right]
\Gamma^2\left[{N\over N+k}\right]}~.
\label{d11}\ee
Now we take the  `t Hooft limit of the correlation function, defined in section \ref{hooft}, to obtain
\be
\Delta_{\Yboxdim4pt\yng(1)}={\l\over 2(1+\l)}\left[1-{1\over N^2}\right]\sp \Delta_A={\l\over 1+\l}~,
\label{d12} \ee
\be
\Delta_A-2\Delta_{\Yboxdim4pt\yng(1)}= {\l\over 1+\l}{1\over N^2}~,
\label{d13}\ee
\be
h={1\over N^2}{\Gamma\left[{\l\over \l+1}\left(1-{1\over N}\right)\right]
\Gamma\left[{\l\over \l+1}\left(1+{1\over N}\right)\right]
\Gamma^2\left[{1\over 1+\l}\right]
\over \Gamma\left[{1+{\l\over N}\over \l+1}\right]\Gamma\left[{1-{\l\over N}\over \l+1}\right]
\Gamma^2\left[{\l\over 1+\l}\right]}={1\over N^2}+{\cal O}\left({1\over N^3}\right)~,
\label{d14} \ee
\be
{\cal F}^{(1)}_1(x)=x^{-{\l\over 1+\l}\left[1-{1\over N^2}\right]}(1-x)^{{\l\over 1+\l}{1\over N^2}}
~F\left({\l\over 1+\l}{1\over N},-{\l\over 1+\l}{1\over N};{1\over 1+\l},x\right)
\label{d15}\ee
$$
=x^{-{\l\over 1+\l}}+{\cal O}\left({1\over N^2}\right)~,
$$
\be
{\cal F}^{(1)}_2(x)={\l\over N}~x^{{1+{\l\over N^2}\over 1+\l}}(1-x)^{{\l\over 1+\l}{1\over N^2}}
~F\left(1+{\l\over 1+\l}{1\over N},1-{\l\over 1+\l}{1\over N};{2+\l\over 1+\l},x\right)
\label{d16} \ee
$$
={\l\over N}~x^{{1\over 1+\l}}~F\left(1,1;{2+\l\over 1+\l},x\right)+{\cal O}\left({1\over N^2}\right)~,
$$
\be
{\cal F}^{(2)}_1(x)=x^{{\l\over 1+\l}{1\over N^2}}(1-x)^{{\l\over 1+\l}{1\over N^2}}
~F\left({\l\over 1+\l}\left(1-{1\over N}\right),{\l\over 1+\l}\left(1+{1\over N}\right);1+{\l\over 1+\l},x\right)
\label{d17}\ee
$$
=~F\left({\l\over 1+\l},{\l\over 1+\l};1+{\l\over 1+\l},x\right)+{\cal O}\left({1\over N^2}\right)
$$
$$
=(1-x)^{1\over 1+\l}~F\left(1,1;1+{\l\over 1+\l},x\right)+{\cal O}\left({1\over N^2}\right)~,
$$
\be
{\cal F}^{(2)}_2(x)=-N~x^{{\l\over 1+\l}{1\over N^2}}(1-x)^{{\l\over 1+\l}{1\over N^2}}
~F\left({\l\over 1+\l}\left(1-{1\over N}\right),{\l\over 1+\l}\left(1+{1\over N}\right);{\l\over 1+\l},x\right)
\label{d18} \ee
$$
=-N~x^{{\l\over 1+\l}{1\over N^2}}(1-x)^{-{\l\over 1+\l}\left[1-{1\over N^2}\right]}
~F\left(-{\l\over 1+\l}{1\over N},{\l\over 1+\l}{1\over N};{\l\over 1+\l},x\right)
$$
$$
=-N~(1-x)^{-{\l\over 1+\l}}+{\cal O}\left({1\over N^2}\right)~.
$$

Using the above the results we find
 \be
G_{11}=|x|^{-{2\l\over 1+\l}}+{\cal O}\left({1\over N^2}\right)\sp  G_{22}=|1-x|^{-{2\l\over 1+\l}}+
{\cal O}\left({1\over N^2}\right)~,
\label{d19} \ee
\be
G_{12}= {\lambda\over N}|x|^{-{2\l\over 1+\l}}~\bar x~F\left(1,1;{2+\l\over 1+\l},\bar x\right)
-{1\over N}(1-x)|1-x|^{-{2\l\over 1+\l}}~F\left(1,1;1+{\l\over 1+\l},x\right)+{\cal O}\left({1\over N^2}\right).
\label{d20} \ee

In the limit $\l\to \infty$ of the leading order in 1/$N$ result we obtain
\be
G_{11}={1\over |x|^{{2}}}+\cdots \sp  G_{22}={1\over |1-x|^{{2}}}+\cdots~,
\label{d21} \ee
\be
G_{12}= {\lambda\over N}{1\over x(1-\bar x)}+\cdots
\label{d22}\ee
where we have used $F(1,1,1,x)={1\over 1-x}$.
This is the free-fermion four-point function.

In the limit $\l\to 0$ we obtain instead
\be
G_{11}=1-2\l \log|x|+\cdots\sp G_{22}=1-2\l\log|1-x|+\cdots
\sp
G_{12}=-{1\over N}+\cdots~.
\label{d23}\ee

\subsection{On the large-$N$ and large-$\l$ limits.}

Finally, we consider the commutativity of the two limits $N\to\infty$ and $\lambda\to\infty$, in one of
the dynamical functions of the WZW model, namely the structure constant $h$, and show that the
two limits do not commute.

We start from (\ref{d11}) and expand the OPE coefficient in a double series
\be
h=\sum_{n=1}^{\infty}\sum_{m=-2}^{\infty}{W_{n,m}\over N^{2n}\l^m}~.
\label{a27b} \ee

Taking the large-$N$ limit first while keeping $\l$ fixed we obtain
\be
\lim_{N\to\infty}~h=\sum_{n=1}^{\infty} {Z_{n}(\l)\over N^{2n}}
\label{a28b} \ee
with
\be
Z_1=1\sp Z_2={\l^2\over (\l+1)^2}\left[\psi'\left({\l\over 1+\l}\right)-\psi'\left({1\over 1+\l}\right)\right]~,
\label{a26}\ee
\be
Z_3={\l^4\over 12(1+\l)^4}\left[6\left[\psi'\left({1\over 1+\l}\right)-\psi'\left({\l\over 1+\l}\right)\right]^2-
\psi'''\left({1\over 1+\l}\right)+\psi'''\left({\l\over 1+\l}\right)\right]~.
\label{a29} \ee
The $\l\to\infty$ limit of these expressions gives
\be
\lim_{\l\to\infty} Z_2=-\l^2+{\cal O}\left({1\over \l}\right)~,
\label{a30a} \ee
\be
\lim_{\l\to\infty} Z_3=2\l\psi''(1)-6\psi''(1)+{\cal O}\left({1\over \l}\right)
\label{a31a} \ee
from which we read in the range $1\leq n\leq 3$, $-2\leq m\leq 0$
\be
W_{1,0}=1\sp W_{2,-2}=-1\sp W_{3,-1}=2\psi''(1)\sp W_{3,0}=-6\psi''(1)
\label{a32} \ee
with all other coefficients being zero in that range.
$\psi(x)=\frac{d \ln \Gamma(x)}{dx}$ is the standard digamma function.

On the other hand, taking the large-$\l$ limit first we obtain
\be
\lim_{\l\to\infty}~h=\sum_{n=-2}^{\infty}H_{n}(N)\l^{-n}
\label{a33a} \ee
with
\be
H_{-2}=-{1\over N^4}\sp H_{-1}={2\over N^4}\left[2\gamma_E
+\psi\left({1\over N}\right)+\psi\left(-{1\over N}\right)\right]~,
\label{a34} \ee
\be
H_0=-{3\psi\left(1+{1\over N}\right)^2+16\gamma_{E}\psi
\left(-{1\over N}\right)+\psi\left(-{1\over N}\right)^2
+16\gamma_{E}\psi\left({1\over N}\right)+\psi\left({1\over N}\right)^2\over 2N^4}-
\label{a35} \ee
$$
-{8\psi\left(1+{1\over N}\right)\psi\left(1-{1\over N}\right)+3\psi\left(1-{1\over N}\right)^2\over 2N^4}
+{2-4\gamma_E(1+2\gamma_E)\over N^4}+
$$
$$
+{1\over N^2}-{\pi\over N^5}\cot\left({\pi\over N}\right)+{2\psi'\left(-{1\over N}\right)-2\psi'\left({1\over N}\right)\over N^5}~.
$$
The large-$N$ limit of these expressions gives
\be
\lim_{N\to\infty} H_{-1}={2\psi''(1)\over N^6}+{\cal O}(N^{-8})~,
\label{a36a} \ee
\be
\lim_{N\to\infty} H_{0}={1-4\gamma_E+{\pi^2\over 3}\over N^4}
+{15\pi^2+\pi^4-180\psi''(1)\over 45N^6}+{\cal O}(N^{-8})
\label{a37a} \ee
from which we deduce
\be
W_{2,-2}=-1\sp W_{3,-1}=2\psi''(1)\sp W_{2,0}=1-4\gamma_E+{\pi^2\over 3}\sp W_{3,0}={15\pi^2+\pi^4-180\psi''(1)\over 45}
\label{a38} \ee
with the rest of the coefficients zero.

Comparing the results (\ref{a32}) and (\ref{a38}) we observe that the two limits do not commute.

\section{YM-CS in AdS$_3$\label{csym}}

\subsection{Conventions, action and gauge invariance}

We work in Euclidean space, with an AdS metric in Poincar\'e coordinates (we set $\ell=1$) with the boundary at $r=0$ being a compact Riemann surface.
We will specialize to the complex plane, with
\be
ds^2={dr^2+(dx^1)^2+(dx^2)^2\over r^2}\sp \e^{r12}=1
\label{ddd1}\ee
Greek indices are three dimensional while latin indices are two dimensional transverse to the boundary. We also use $\e^{ij}=\e^{rij}$.

It will be eventually convenient to define complex coordinates
\be
z={x^1+ix^2\over \sqrt{2}}\sp x^1={z+\bz\over \sqrt{2}}\sp x^2={z-\bz\over i\sqrt{2}}
\label{dd5}\ee
and for the gauge field
\be
A_1={A_z+A_{\bz}\over \sqrt{2}}\sp A_2={A_z-A_{\bz}\over i\sqrt{2}}
\label{dd6}\ee
The metric becomes
\be
ds^2={dr^2+2dzd\bz\over r^2}\sp \e^{rz\bz}=i
\label{dd7}\ee

The action has bulk components and boundary ``counter terms".
\be
S=S_{\rm bulk}+S_{\rm ct}\sp S_{\rm bulk}=S_{CS}+S_{YM}
\label{dd2}\ee
\be
S_{CS}={ik\over 16\pi}\int_{\cal M} d^3x ~\e^{\m\n\r}A_{\m}F_{\n\r}\sp S_{YM}={1\over 4g^2}\int_{\cM} d^3x~\sqrt{g}~F_{\m\n}F^{\m\n}
\label{dd3}\ee
\be
S_{ct}=-{|k|\over 16\pi}\int_{\cdM}d^2x \sqrt{g}~g^{ij}A_{i}A_{j}
\label{dd4}\ee
We have specialized to the abelian case, and in our normalization $k\in Z$.

There is another possible counterterm $\int_{\cdM}d^2x \sqrt{g}g^{\m\n}A_{\m}F_{r\n}$, but we will see later that this will vanishes identically
on the solutions.
We also define the sign $\xi$ as
\be
{k\over |k|}=\xi
\label{dd4a}\ee
and define the projector
\be
P_{\xi}^{ij}\equiv g^{ij}+i\xi {\e^{ij}\over \sqrt{g}}
\ee

Under a U(1) gauge transformation
\be
A_{\m}\to A_{\m}+\pa_{\m}\Lambda
\label{dd8}\ee
the action transforms as follows
\be
\delta_{\Lambda} S_{YM}=0\sp \delta_{\Lambda} S_{CS}={ik\over 16\pi}\int_{\cdM}d^2x~\e^{ij}~\Lambda F_{ij}\sp \delta_{\Lambda}
 S_{ct}={|k|\over 8\pi}\int_{\cdM}d^2x \sqrt{\hat g}~\hat g^{ij}~\Lambda ~\pa_{i}A_{j}
\label{dd9}\ee
where $\hat g_{ij}=g_{ij}\Big |_{r=0}$ is the boundary metric.

Putting everything together we obtain
\be
\delta_{\Lambda} S={|k|\over 8\pi}\int_{\cdM}d^2x\sqrt{\hat g}~\Lambda~\hat P^{ij}_{\xi}~\pa_iA_j=
\left\{ \begin{array}{ll}
\displaystyle {|k|\over 4\pi}\int_{\cdM}d^2x ~\Lambda~\pa_{z}A_{\bz} \;\;\;&\;\;\;\xi=-1,\\
\\
\displaystyle {|k|\over 4\pi}\int_{\cdM}d^2x ~\Lambda~\pa_{\bz}A_{z} \;\;\;&\;\;\;\xi=1.
\end{array}\right.
\label{dd10}\ee

\subsection{Equations of motion}
The relevant variations leading to the equations of motion are
\be
\delta S_{CS}={ik\over 8\pi}\int_{\cM} d^3x~\e^{\m\n\r}\delta A_{\m}F_{\n\r}-{ik\over 8\pi}\int_{\cdM} d^2x~\e^{ij}A_{i}\delta A_{j}
\label{dd11}\ee
\be
\delta S_{YM}={1\over g^2}\int_{\cM}d^3x~\da_{\m}\pa_{\n}\left(\sqrt{g}g^{\m\a}g^{\n\b}F_{\a\b}\right)+{1\over g^2}\int_{\cdM}d^2x~\sqrt{g}g^{rr}g^{ij}F_{ri}\da_{j}
\label{dd12}\ee
\be
\delta S_{ct}=-{|k|\over 8\pi}\int_{\cdM}d^2x \sqrt{g}~g^{ij}A_{i}\da_{j}
\label{dd13}\ee
In total we have
\be
\delta S=\delta S_{3d}+\delta S_{2d}
\label{dd14}\ee
\be
\delta S_{3d}=\int_{\cM}d^3x\left[{1\over g^2}\pa_{\n}\left(\sqrt{g}g^{\m\a}g^{\n\b}F_{\a\b}\right)+{ik\over 8\pi}\e^{\m\n\r}F_{\n\r}\right]~\da_{\m}
\label{dd15}\ee
\be
\delta S_{2d}=\int_{\cdM}d^2x~\sqrt{g}~\left[{|k|\over 8\pi}\hat P^{ij}_{\xi}A_i+
{1\over g^2}g^{rr}g^{ij}F_{ri}\right]\da_j
\label{dd16}\ee
\be
=
\left\{ \begin{array}{ll}
\displaystyle -{|k|\over 4\pi}\int_{\cdM}d^2x ~A_{z}\delta A_{\bz}+{1\over g^2}\int_{\cdM}d^2x~\sqrt{g}~g^{rr}g^{ij}F_{ri}\da_j \;\;\;&\;\;\;\xi=-1,\\
\\
\displaystyle -{|k|\over 4\pi}\int_{\cdM}d^2x ~A_{\bz}\delta A_{z}+{1\over g^2}\int_{\cdM}d^2x~\sqrt{g}~g^{rr}g^{ij}F_{ri}\da_j \;\;\;&\;\;\;\xi=+1.
\end{array}\right.
\label{dd16a}
\ee
It will be shown later that the YM boundary terms vanish at the boundary.
From (\ref{dd16a})  we also obtain the boundary currents
\be
J_z\equiv -2\pi{\delta S\over \delta A_{\bz}}=0\sp J_{\bz}\equiv -2\pi{\delta S\over \delta A_{z}}={|k|\over 2}A_{\bz}\sp \xi=1
\label{dd16b}\ee
\be
J_z\equiv -2\pi{\delta S\over \delta A_{\bz}}={|k|\over 2}A_{z}\sp J_{\bz}\equiv -2\pi{\delta S\over \delta A_{z}}=0\sp \xi=-1
\label{dd16c}\ee

From (\ref{dd15}) follow the bulk equations of motion
\be
\pa_{\n}\left(\sqrt{g}g^{\m\a}g^{\n\b}F_{\a\b}\right)+{ig^2k\over 8\pi}\e^{\m\n\r}F_{\n\r}=0
\label{dd17}\ee
which describe a massive U(1) gauge field with mass $|m|$ where
\be
m={g^2k\over 4\pi}
\ee
In components they read
\be
r\left(\pa_{z}F_{r\bz}+\pa_{\bz}F_{rz}\right)-mF_{z\bz}=0
\label{dd18}\ee
\be
\pa_r(rF_{r\bz})+r\pa_{\bz}F_{z\bz}-mF_{r\bz}=0
\label{dd19}\ee
\be
\pa_r(rF_{rz})-r\pa_{z}F_{z\bz}+mF_{rz}=0
\label{dd20}\ee

We now fix the gauge $A_r=0$, which implies $F_{rz}=\pa_r A_{z}$, $F_{r\bz}=\pa_{r}A_{\bz}$.
Equations (\ref{dd18})-(\ref{dd20}) become
\be
r\pa_r\left(\pa_{z}A_{\bz}+\pa_{\bz}A_{z}\right)-mF_{z\bz}=0
\label{dd18a}\ee
\be
{1\over r}\pa_r(r\pa_{r}A_{\bz})+\pa_{\bz}F_{z\bz}-{m\over r}\pa_rA_{\bz}=0
\label{dd19a}\ee
\be
{1\over r}\pa_r(r\pa_r A_{z})-\pa_{z}F_{z\bz}+{m\over r}\pa_r A_{z}=0
\label{dd20a}\ee

The above equations have step two in $r$, so we expand in a power series in $r^2$
\be
A_{z}=r^a\sum_{n=0}^{\infty} A^{(n)}_{z}r^{2n}\sp A_{z}=r^{\bar a}\sum_{n=0}^{\infty} A^{(n)}_{\bz}r^{2n}
\label{dd21}\ee
We will assume $a=\bar a$ in the sequel as in the opposite case we do not seem to obtain non-trivial solutions.

From (\ref{dd18a}) we obtain
\be
(2n+ a)(\pa_z  A^{(n)}_{\bz}+\pa_{\bz} A^{(n)}_{z})=mF^{(n)}_{z\bz}
\label{dd22}\ee
From (\ref{dd19a}), (\ref{dd20a}) we obtain
\be
(2n+a+2)\left(2n+a+2-m\right)A_{\bz}^{(n+1)}+\pa_{\bz}F^{(n)}_{z\bz}=0\sp a\left(a-m\right)A_{\bz}^{(0)}=0
\label{dd23}\ee
\be
(2n+a+2)\left(2n+a+2+m\right)A_{z}^{(n+1)}-\pa_{z}F^{(n)}_{z\bz}=0\sp a\left(a+m\right)A_{z}^{(0)}=0
\label{dd24}\ee

The indicial equations are satisfied for $a=0$. We will investigate later the possibility that $a\not =0$.
Then (\ref{dd22}) implies that
\be
F^{(0)}_{z\bz}=0
\label{dd25}\ee

and (\ref{dd23}), (\ref{dd24}) become
\be
4(n+1)\left(n+1-{g^2k\over 8\pi}\right)A_{\bz}^{(n+1)}+\pa_{\bz}F^{(n)}_{z\bz}=0
\label{dd23a}\ee
\be
4(n+1)\left(n+1+{g^2k\over 8\pi}\right)A_{z}^{(n+1)}-\pa_{z}F^{(n)}_{z\bz}=0
\label{dd24a}\ee

Note that from (\ref{dd9}), because of (\ref{dd25}), the bulk action is {\it gauge invariant}.

We now rewrite (\ref{dd22}) as
\be
\left(n-{m\over 2}\right)\pa_zA_{\bz}^{(n)}+\left(n+{m\over 2}\right)\pa_{\bz}A^{(n)}_{z}=0
\label{dd26}\ee
and use it to rewrite (\ref{dd23a}) and (\ref{dd24a}) as
\be
A_{\bz}^{(n+1)}=-{n\over 2(n+1)\left(n+1-{m\over 2}\right)\left(n+{m\over 2}\right)}\pa_z\pa_{\bz}A^{(n)}_{\bz}
\label{dd27}
\ee
\be
A_{z}^{(n+1)}=-{n\over 2(n+1)\left(n+1+{m\over 2}\right)\left(n-{m\over 2}\right)}\pa_z\pa_{\bz}A^{(n)}_{z}
\label{dd28}
\ee
Notice that the formulae above work as long as ${m\over 2}$ is not an integer.
From (\ref{dd23}), (\ref{dd24}) we obtain
\be
4\left(1-{m\over 2}\right)A^{(1)}_{\bz}=-\pa_{\bz}F^{(0)}_{z\bz}=0
\sp 4\left(1+{m\over 2}\right)A^{(1)}_{z}=\pa_{z}F^{(0)}_{z\bz}=0
\label{dd29}\ee

We will now consider first the generic case, $m$ real.
We obtain $A^{(m>0)}_{z,\bz}=0$, a trivial solution.
We are left with $A^{(0)}_{z,\bz}$ satisfying $F^{(0)}_{z\bz}=0$.
Therefore, the source can be a flat boundary field only.

\subsubsection{The case $m=\pm 2$}

In this case
\be
A_{\bz}^{(n>0)}={(-1)^{n+1}\over 2^{n-1}(n!)^2}(\pa_{z}\pa_{\bz})^{n-1}A_{\bz}^{(1)}
\sp A_{z}^{(n>0)}=0\sp m= 2
\label{dd30}\ee
and $A^{(1)}_{\bz}$ still remains arbitrary.
The only $z$ component that remains is $A_{z}^{(0)}$.
 This is not compatible with the boundary variation (\ref{dd16a}) as here $k>0$ and $\xi=1$ except if it vanishes.

 We can resum $A_{\bz}$ as follows
 \be
 A_{\bz}(r,z,\bz)=A_{\bz}^{(0)}(z,\bz)+G\left(-{1\over 2}\pa_{z}\pa_{\bz}\right)A_{\bz}^{(1)}(z,\bz)
\label{dd33}\ee
with
\be
G(x)={1\over x}\left[\int_1^{x}{e^u\over u}du+1-\log x\right]
~.
\label{dd34}\ee

In the opposite case
\be
A_{\bz}^{(n>0)}={(-1)^{n+1}\over 2^{n-1}(n!)^2}(\pa_{z}\pa_{\bz})^{n-1}A_{z}^{(1)}
\sp A_{\bz}^{(n>0)}=0\sp m=- 2
\label{dd30a}\ee
and $A^{(1)}_{z}$ still remains arbitrary.
In this case the only $\bz$ component that remains is $A_{\bz}^{(0)}$ satisfying $F_{z\bz}^{(0)}=0$.
This is incompatible with the boundary variation (\ref{dd16a}) as here $k<0$ and $\xi=-1$ except if it vanishes.

To summarize the above discussion.

1) If $k>0$ and $m=2$, the general solution has  $A^{(n\not =0)}_z=0$ for all $n$ and $ A^{(0)}_{\bz}(z)\not=0$, $A^{(1)}_{\bz}(z,\bz)\not=0$, the higher $A_{\bz}^{(n>1)}=0$.
It gives nontrivial $J_{\bz}$ currents on the boundary if we set $A^{(0)}_z=0$.

2) If $k<0$ and $m=-2$, the general solution has  $A^{(n\not =0)}_{\bz}=0$ for all $n$ and $ A^{(0)}_{z}(\bz)\not=0$, $A^{(1)}_{z}(z,\bz)\not=0$ the higher $A_{z}^{(n>1)}=0$.
It gives nontrivial $J_{z}$ currents on the boundary if we set $A^{(0)}_{\bz}=0$.

\subsubsection{The case $m=\pm 2 N$, $N=2,3,\cdots$}
If  $m=2N>0$
\be
A_{\bz}^{(n+N)}={(-1)^{n}N!\over 2^{n}(n!)^2(N+n)}{n\choose{n+2N-1}}^{-1}(\pa_{z}\pa_{\bz})^{n}A_{\bz}^{(N)}
\sp A_{z}^{(n>0)}=0
\label{dd301}\ee
\be
A^{(n)}_{\bz}=0\;\;,\;\;n=1,2,,\cdots,N-1
\label{dd35}\ee
and $A^{(N)}_{\bz}$ still remains arbitrary.

In this case the only $z$ component that remains is again $A_{z}^{(0)}$.

In the opposite case,  $m=-2N<0$.
\be
A_{\bz}^{(n+N)}={(-1)^{n+}N!\over 2^{n}(n!)^2(N+n)}{n\choose{n+2N-1}}^{-1}(\pa_{z}\pa_{\bz})^{n}A_{z}^{(N)}
\sp A_{\bz}^{(n>0)}=0
\label{dd30ab}\ee
and $A^{(1)}_{z}$ still remains arbitrary.

In all the cases above $F_{rz},F_{r\bz}$ vanishes at the boundary.

\subsection{The on-shell action}

We rewrite $S_{bulk}$ as
\be
S_{bulk}=\int_{\cM}d^3x~A_{\m}\left[{1\over 2g^2}\pa_{\n}\left(\sqrt{g}g^{m\a}g^{\n\b}F_{\a\b}\right)+{ik\over 16\pi}\e^{\m\n\r}F_{\n\r}\right]+{1\over 2g^2}\int_{\cM}d^3x
\pa_{\m}\left(\sqrt{g}g^{m\a}g^{\n\b}A_{\n}F_{\a\b}\right)
\label{dd31}\ee
where we separated the first part that is proportional to the bulk equations of motion, and will vanish on shell.

We obtain the on-shell action $S_{\rm on-shell}=S_{\rm bulk}+S_{ct}$ as
\be
S_{\rm on-shell}={1\over 2g^2}\int_{\cdM}d^2x~r\left(A_zF_{r\bz}+A_{\bz}F_{rz}\right)-{|k|\over 8\pi}\int_{\cdM}d^2x ~A_{z}A_{\bz}
\label{dd32}\ee
$$
=-{|k|\over 8\pi}\int_{\cdM}d^2x ~A_{z}A_{\bz}
$$
as the first contributions vanishes at the boundary.
This is the full answer for the abelian theory.
In the non-abelian theory the boundary term generated by the YM action will also vanish.
In the CS action what will remain is the cubic term as described in the main part of this paper.

Note that if we generalize the YM action to contain any non-linear function of the field strength and its derivatives, all its contributions on-shell will give boundary terms
that will vanish.

\subsection{Equations of motion in momentum space}

In Cartesian coordinates $r,x^{1,2}$ the equations are
\be
r(\pa_1F_{r1}+\pa_2F_{r2})+{im }F_{12}=0
\label{ddb1}\ee
\be
-\pa_r(rF_{r1})+\pa_2(rF_{12})-{im }F_{r2}=0
\label{ddb2}\ee
\be
\pa_r(rF_{r2})+\pa_1(rF_{12})-{im }F_{r1}=0
\label{ddb3}\ee

We Fourier transform $\pa_i\to ik_i$ to obtain
\label{ddb4}\be
ir\pa_r(k_1A_1+k_2A_2)+r(k_1^2+k_2^2)A_r-{m }(k_1A_2-k_2A_1)=0
\label{ddb5}\ee
\be
-\pa_r(r\pa_rA_1)+rk_2^2 A_1-rk_1k_2A_2-{im }\pa_r A_2+ik_1\pa_r(rA_r)-{m }k_2A_r=0
\label{ddb6}\ee
\be
\pa_r(r\pa_rA_2)-rk_1^2A_2+rk_1k_2A_1-{im }\pa_rA_1-ik_2\pa_r(rA_r)-{m }k_1A_{r}=0
\label{ddb7}\ee
We now decompose $A_i=A^{\perp}_{i}+ik_i\phi$, with $k\cdot A^{\perp}=0$ and $\vec k^2\not= 0$ to obtain,

\be
r\vec k^2(A_r-\pa_r\phi)-{m }(k_1A^{\perp}_2-k_2A^{\perp}_1)=0
\label{ddb8}\ee
\be
\pa_r(r\pa_rA^{\perp}_1)-rk_2^2 A^{\perp}_1+rk_1k_2A^{\perp}_2+{im }\pa_r A^{\perp}_2-ik_1\pa_r(r(A_r-\pa_r\phi))+{m}k_2(A_r-\pa_r\phi)=0
\label{ddb9}\ee
\be
\pa_r(r\pa_rA^{\perp}_2)-rk_1^2A^{\perp}_2+rk_1k_2A^{\perp}_1-{im }\pa_rA^{\perp}_1-ik_2\pa_r(r(A_r-\pa_r\phi))-{m}k_1(A_{r}-\pa_r\phi)=0
\label{ddb10}\ee

Solving $k\cdot A^{\perp}=0$ and substituting in (\ref{ddb8}) we obtain
\be
A_r-\pa_r\phi=-{m\over k_2 r}A^{\perp}_1
\ee
Substituting this in (\ref{ddb9}) and (\ref{ddb10}) we obtain
\be
{1\over r}\pa_r(r\pa_r A_i^{\perp})
-\left(\vec k^2 +{m^2\over r^2}\right)A_i^{\perp}=0
\ee
This is the Bessel equation and the regular solution is
\be
 A_i^{\perp}=C_i~K_{|m|}(|\vec k| r)\sp \vec k\cdot \vec C=0
 \ee
Note that as $K_{\nu}\sim r^{-|\nu|}$ for $r\to 0$, this is singular at the boundary.

For the constant mode $k_1=k_2=0$ we obtain instead as the only non-trivial equations

\be
\pa_r(r\pa_rA_1)+im\pa_rA_2=0\sp \pa_r(r\pa_rA_2)-im\pa_rA_1=0
\ee
with general solution
\be
A_1=C_+r^{|m|}+C_- r^{-|m|}+C_1\sp A_2=i{m\over |m|}(C_+r^{|m|}-C_-r^{-|m|})+C_2
\ee

Both of  these solutions are {\it singular} at the boundary.

\subsection{Chern-Simons action only}

We now consider the non-abelian action
\be
S_{bulk}={ik\over 8\pi}Tr\int_{\cM} d^3x \epsilon^{\m\n\r}\left(A_{\m} \pa_{\n}A_{\r}+{2\over 3}A_{\mu}A_{\nu}A_{\r}\right)
\label{dd38}\ee
$$
={ik\over 16\pi}Tr\int_{\cM} d^3x \epsilon^{\m\n\r}\left(A_{\m} F_{\n\r}-{2\over 3}A_{\mu}A_{\nu}A_{\r}\right)
$$
The bulk equations of motion are obtained
from the variation
\be
\delta S_{bulk}={ik\over 8\pi}Tr\int_{\cM} d^3x \epsilon^{\m\n\r}~\delta A_{\m}~F_{\n\r}-{ik\over 8\pi}\int_{\cdM}d^2x \e^{ij}A_i\delta A_j
\label{dd39}\ee
and are the flatness condition
\be
F_{\m\n}=\pa_{\m}A_{\n}-\pa_{\n}A_{\m}+[A_{\m},A_{\n}]=0
\label{dd40}\ee
with general solution
\be
A_{\m}=U^{-1}\pa_{\mu}U\sp \epsilon^{\mu\nu\r}\pa_{\m}A_{\n}=-\epsilon^{\mu\nu\r}A_{\mu}A_{\n}
\label{dd41}\ee
where $U(r,z,\bz)$ is a group element.
Therefore
\be
S_{\rm bulk}^{\rm on-shell}=-{ik\over 24\pi}Tr\int_{\cM} d^3x \epsilon^{\m\n\r}U^{-1}\pa_{\mu}UU^{-1}\pa_{\nu}UU^{-1}\pa_{\r}U=k\Gamma(U)
\label{dd42}\ee
where $\Gamma(U)$ is the properly normalized WZ term.
The counterterm is
\be
S_{ct}=-{|k|\over 16\pi}Tr\int_{\cdM}d^2x \sqrt{g}~g^{ij}A_{i}A_{j}
\label{dd43}\ee
with
\be
\delta S_{ct}=-{|k|\over 8\pi}Tr\int_{\cdM}d^2x \sqrt{g}~g^{ij}A_{i}\delta A_{j}
\label{dd44}\ee
so that
\be
\delta(S_{bulk}+S_{ct})={ik\over 8\pi}Tr\int_{\cM} d^3x \epsilon^{\m\n\r}~\delta A_{\m}~F_{\n\r}-{|k|\over 8\pi}Tr\int_{\cdM}d^2x \sqrt{g}~P^{ij}_{\xi}A_{i}\delta A_j
\label{dd45}\ee
Therefore,
\be
S_{ct}^{on-shell}={|k|\over 16\pi}Tr\int d^2x \sqrt{\hat g}\hat g^{ij}\pa_iU\pa_j U^{-1}
\label{dd46}\ee
In total
\be
S^{on-shell}=S_{WZW}(U)
\label{dd47}\ee
Under gauge transformations
\be
A_{\m}'=U(A_{\m}-U^{-1}\pa_{\m}U)U^{-1}\sp F'_{\m\n}=UF_{\m\n}U^{-1}
\label{dd48}\ee
\be
\delta A_{\m}=A'_{\m}-A_{\m}=D_{\m}\e\equiv\pa_{\m}\e+[A_{\m},\e]\sp \delta F_{\m\n}=[F_{\m\n},\e]
\label{dd49}\ee
and
\be
S'_{\rm bulk}(A')=S_{\rm bulk}(A)-{ik\over 2\pi}Tr\int_{\cM} d^3x \epsilon^{\m\n\r}\left(U^{-1}\pa_{\m}UU^{-1}\pa_{\n}UA_{\r}-{1\over 6}U^{-1}\pa_{\m}UU^{-1}\pa_{\n}UU^{-1}\pa_{\r}U\right)
\label{dd50}\ee
$$
+{ik\over 4\pi}Tr\int d^2x \e^{ij}U^{-1}\pa_{i}U~A_j
~.
$$

The infinitesimal variation becomes using (\ref{dd39})
\be
\delta_{\e}S_{\rm bulk}={ik\over 8\pi}Tr\int d^3x\e^{\m\n\r}~\e D_{\mu}F_{\n\r}-{ik\over 8\pi}\int_{\cdM}d^2x~\e^{ij}~Tr\left[A_iD_j\e-\e F_{ij}\right]
\ee
$$
={ik\over 8\pi}\int_{\cdM}d^2x~\e^{ij}~\e\pa_iA_j
$$
where we have used
\be
\e^{\m\n\r}D_{\m}F_{\n\r}=\e^{\m\n\r}\left(\pa_{\m}F_{\n\r}+[A_{\m},F_{\n\r}]\right)=0
\ee

The formulae above indicate that we can make gauge transformations that are trivial on the boundary. If we transform by the group element $V(r,z,\bz)$ we obtain
\be
A_{\mu}=(UV)^{-1}\pa_{\mu}(UV)
\ee
with
\be
U(r,z,\bz)=U^{(0)}(z,\bz)X(r,z,\bz)\sp X(r,z,\bz)=1+{\cal O}(r)\sp V(r,z,\bz)=1+ {\cal O}(r)
\ee
we may choose $V=X^{-1}$ to obtain
\be
A_r=0\sp A_{i}=(U^{(0)})^{-1}\pa_iU^{(0)}\sp F^{(0)}_{ij}=0
\ee

In this case, all the $r$ dependence is a gauge artifact and can be eliminated.

\end{appendix}

\end{document}